# An adaptive time-tree transition kernel for Bayesian phylogenetic inference

Marius Brusselmans[1,*] Guy Baele[1] Samuel L. Hong[1] Jiansi Gao[2]
Marc A. Suchard[3,4,5] Andrew Rambaut[6] Luiz Max Carvalho[6,7]

[1]Department of Microbiology, Immunology and Transplantation, Rega Institute, KU Leuven, Leuven, Belgium
[2]Computational Biology Program, Fred Hutchinson Cancer Center, Seattle, Washington, United States
[3]Department of Biostatistics, School of Public Health, University of California, Los Angeles, United States
[4]Department of Biomathematics, David Geffen School of Medicine at UCLA, University of California, Los Angeles, United States
[5]Department of Human Genetics, David Geffen School of Medicine at UCLA, Universtiy of California, Los Angeles, United States
[6]Institute of Ecology and Evolution, University of Edinburgh, Edinburgh, United Kingdom
[7]School of Applied Mathematics, Getulio Vargas Foundation, Rio de Janeiro - RJ, Brazil

[*]Corresponding author: Rega Institute, Herestraat 49, 3000 Leuven, Belgium. E-mail: marius.brusselmans@kuleuven.be

**Abstract**

Bayesian phylogenetic and phylodynamic analyses can be very time-consuming, owing to the combination of complex models that are used to estimate key parameters from increasingly large genomic data sets and their associated metadata. The use of high-performance computer hardware can – to a certain extent – alleviate the computational burden and markedly decrease the time to results. Still, even converging to the posterior can be a lengthy endeavour, with the burn-in aspect of such analyses potentially taking days or even weeks for large data sets. One of the key aspects that hampers performance in Bayesian phylogenetic inference is the efficiency with which tree topology proposals explore tree space. We here propose a novel adaptive tree transition kernel, which we call 'subTreeLeap' (STL), which involves modifying the phylogeny by walking along patristic distance paths in the tree according to an adaptable radius parameter. STL is a general proposal, which can be used with contemporaneous or time-calibrated sequence data, being particularly suited to the latter due to respecting temporal precedence constraints. We carefully assess its impact on convergence and statistical mixing of the exploration of posterior tree space, by comparison to replicate "golden runs" obtained from lengthy analyses of empirical data under standard tree transition kernels. We find that STL successfully explores the same posterior tree space as standard kernels, but often does so in a more efficient manner. We discuss limitations as well as future potential improvements to STL that could substantially increase the speed at which Bayesian phylogenetic inferences are obtained.

# Introduction

In Bayesian phylogenetics, one is usually interested in computing the posterior distribution $p(t, \boldsymbol{b}, \boldsymbol{\theta}|D)$, where $D$ represents the observed data, $\boldsymbol{T}$ the set of all binary rooted trees, $t \in \boldsymbol{T}$ a tree topology associated with a set of branch lengths $\boldsymbol{b} = \boldsymbol{B}(t) = \{b_1, b_2, \ldots, b_{2n-2}\}$, and $\boldsymbol{\theta}$ a set of parameters of interest such as substitution model parameters, molecular clock model parameters, migration rates, heritability coefficients, etc. One of the main features of the Bayesian approach is to enable parameter inference and hypothesis testing whilst accommodating phylogenetic uncertainty (Suchard et al., 2001; Huelsenbeck et al., 2002).

This treatment of uncertainty is achieved by integrating over the space of phylogenetic trees, which crucially depends on efficiently traversing tree space. Even for the simplest of models, the posterior distribution cannot be computed analytically and requires numerical approximation, usually accomplished through Markov chain Monte Carlo (MCMC). The use of MCMC for Bayesian methods in phylogenetics has grown steadily since its introduction in the late 1990s and early 2000s (Sinsheimer et al., 1996; Yang and Rannala, 1997; Mau et al., 1999; Li et al., 2000; Huelsenbeck et al., 2001), with software packages such as MrBayes (Ronquist et al., 2012) and BEAST (Suchard et al., 2018; Bouckaert et al., 2019; Baele et al., 2025) becoming widely used by researchers in a broad range of disciplines (Murphy et al., 2001; Bouckaert et al., 2012; Lemey et al., 2014). In many applications, interest lies in constructing time-calibrated phylogenies, i.e. phylogenetic trees whose branch lengths are measured in units of calendar time. In particular, one might have sequences sampled through time (heterochronous) which enable direct estimation of the rate of evolution and reconstruction of past population dynamics (Drummond et al., 2002, 2005). Populations for which such analyses are possible are known as measurably evolving populations (MEPs) (Drummond et al., 2003). These types of MEP data sets pose additional challenges to inference because they impose constraints on the space of valid trees (Stadler and Yang, 2013).

The Metropolis-Hastings (MH) algorithm (Metropolis et al., 1953; Hastings, 1970) is a very popular MCMC technique due to its generality and ease of implementation. In MH, a Markov chain is constructed such that its limiting distribution is the desired posterior distribution. For ease of presentation, let $\tau = (t, \boldsymbol{b})$ be a phylogeny with topology $t$ and branch lengths $\boldsymbol{b}$, and $q_\gamma(\tau' \mid \tau)$ be a conditional distribution indexed by a parameter $\gamma$, from which a new state $\tau'$ can be proposed from the current state $\tau$, which we call a **tree transition kernel**. For a suitably constructed tree transition kernel $q_\gamma$, it can be shown that accepting/rejecting a new state $\tau'$ based on the acceptance ratio $A_\gamma(\tau \mid \tau') = \min\left(1, \frac{p(\tau'|D)q_\gamma(\tau|\tau')}{p(\tau|D)q_\gamma(\tau'|\tau)}\right)$ leads to a Markov chain which has the desired (target) distribution $p(\cdot \mid D)$ as its invariant distribution.

An important aspect of MCMC implementation is that there are no "default" choices for the proposal distribution $q_\gamma(\cdot|\cdot)$; it must be chosen with the target (posterior) distribution in mind. Moreover, the efficiency of MCMC algorithms in approximating the target distribution depends crucially on the choice of transition kernel (Brooks et al., 2003; Al-Awadhi et al., 2004; Yang and Rodríguez, 2013). As argued by Höhna and Drummond (2012), tree transition kernels are usually built in a relatively simplistic fashion, which in turn leads to inefficient exploration of tree space. Most tree transition kernels proposed to date are not adaptive, i.e., the parameter(s) $\gamma$ cannot be adjusted during the Markov chain to achieve a desired acceptance probability. Given the clear advantage of adaptive MCMC over non-adaptive implementations for high-dimensional target distributions (Roberts and Rosenthal, 2009; Baele et al., 2017; Meyer, 2021), the development of adaptive tree transition kernels could lead to substantial gains in exploring posterior tree space.

Lakner et al. (2008) were amongst the first to systematically investigate tree transition kernel efficiency in MCMC for Bayesian phylogenetics. They investigated the performance of seven kernels on a collection of 10 real-world data sets, the now famous 'DS' data sets. To quantify performance, the authors looked at the percentage of converged runs per tested kernel, using clade frequencies relative to a reference (golden) run as a criterion. In addition, time to convergence was also used as a performance criterion. Höhna et al. (2008) developed new "clock-constrained" tree transition kernels to improve convergence and mixing efficiency when estimating time-calibrated trees. The authors argue that clock-constrained trees impose additional restrictions on the state space of the MCMC algorithm and hence that performance could be increased by developing tree transition kernels that took the extra information provided by tip dates. They develop two such kernels: fixed node-height prune-and-regraft (FNPR) and intermediate exchange (IE). FNPR finds a "target" node (excluding the root and its two daughters) at random, prunes it and regrafts the resulting subtree at a "destination" node in the tree at the same height at random. IE is similar in spirit to FNPR, but the regraft node is not chosen uniformly; instead, IE is constructed to prefer local rearrangements, by picking closer nodes with a higher probability – see below and the Methods section in Höhna et al. (2008) for details. A limitation of FNPR and IE though is that they are not adaptive.

Höhna and Drummond (2012) explored more sophisticated "guided" tree transition kernels, inspired by Gibbs sampling. The idea behind their "metropolised" Gibbs samplers is to maximise transition probability, i.e., the probability that the chain moves to a new state. This is accomplished by prohibiting the current state as a proposed state, leading to a transition probability of 1. The transition kernels developed in Höhna and Drummond (2012) use a weighting scheme based on conditional clade probabilities (CCP) to guide transitions between trees. A move to a tree with a lower CCP score is thus less likely, whilst a move that increases the score has a higher probability of being accepted. A limitation of these metropolised transition kernels is that CCP scores require normalisation over all trees (Larget, 2013) and hence can be cumbersome to calculate. More recently, Meyer (2021) investigated a class of adaptive proposals for unrooted topologies which also rely on split frequencies, which are then estimated using a heuristic that ensures stability in the estimates. The author also focuses on solely modifying the topology $t$, relying on mapping techniques or surrogate proposals to update branch lengths. This might lead to poor performance, as we expect the posterior distribution of certain branches to be strongly dependent on the topology in which a given branch exists. Finally, recent work on parsimony-informed proposals (Zhang et al., 2020; Bouckaert et al., 2025; Varilly et al., 2025) has shown these approaches to improve mixing in certain regimes, such as there being few mutations per branch.

To achieve maximum efficiency, a tree transition kernel needs to have the following qualities: (i) be computationally cheap; (ii) be adaptive; (iii) traverse tree space efficiently, i.e. lead to a high level of mixing. We here develop and assess the performance of a new adaptive time-tree transition kernel, which we implement in the open source software package BEAST X (v1.10.5) (Baele et al., 2025). Our main contributions are twofold: we propose a new tree transition kernel that simultaneously proposes changes to topology and branch lengths, and we employ state-of-the-art validation and diagnostic tools in order to assess correctness and efficiency in real-world data sets, focusing on the inference of time-calibrated phylogenies.

# Methods

## Preliminaries

We will here use $\boldsymbol{\Psi} \equiv \mathbb{T}_n \times \boldsymbol{B}$ to denote the parameter space encompassing topologies and branch lengths, henceforth called "tree space", and $\tau \in \boldsymbol{\Psi}$ to denote a bifurcating, rooted tree with branch lengths on $n$ taxa. Let $\mathrm{H}(t) := \boldsymbol{h} = \{h_1, h_2, \ldots, h_{2n-1}\}$ be the set of node heights for all nodes (internal and external) in $\tau$. It is also convenient to define $p_i$, $g_i$ and $s_i$ as the parent, grandparent and sibling of node $i \in \tau$, respectively. $MRCA_{i,j}$ is the most recent common ancestor of nodes $i$ and $j$. Finally, let $\Delta_{ij} = 2h(\mathrm{MRCA}_{ij}) - (h(i) + h(j))$ be the patristic (path) distance between nodes $i$ and $j$ on the phylogeny.

## Tree transition kernels

We here focus on time-calibrated phylogenies (also called time trees), in which branch lengths are measured in units of calendar time and which often present temporal precedence constraints which necessitate the development of special conditional distributions. In guise of example, consider the Fixed Node-height Prune-and-Regraft (FNPR) and Intermediate Exchange (IE) transition kernel developed by Höhna et al. (2008). FNPR regrafts the pruned node $i$ at the same height in a random location in the phylogeny. IE meanwhile exchanges the pruned node $i$ with a randomly selected existing node $j$, assigning probabilities to destinations based on their patristic proximity to $i$: $P_i(j) = \Delta_{ij}/\sum_{k=1}^{n_i} \Delta_{ik}$, where $n_i$ is the number of nodes that can be a regraft point for $i$, i.e. that meet the height constraint.

The first new transition kernel we propose, *SubTreeJump* (STJ), combines elements of both Intermediate Exchange and FNPR. Like FNPR, STJ regrafts the pruned node $i$ at the same height somewhere else in the phylogeny. The different candidate destination nodes are assigned probabilities based on their patristic proximity to $i$, similar to IE. STJ further extends this notion by introducing a tuning parameter that controls how local the proposed rearrangements are. The idea is to make the probability of moving node $i$ to $j$ proportional to the normalized patristic distance between the nodes raised to some power $\alpha$. Given that the normalized patristic distance will be bounded by 0 and 1, this allows one to favour bold moves away from $i$ ($\alpha < 0$) or more conservative moves closer to $i$ ($\alpha > 0$). STJ is thus a generalization of FNPR when $\alpha = 0$, removing the influence of patristic distances. One can then define the STJ transition kernel as $q_\alpha(\tau'|\tau) = \Pr(i \to j)$. The necessary steps to perform a STJ operation on $\tau$ in order to propose a tree $\tau'$ are described in Algorithm 1; we also provide an illustration in Figure 1.

**Algorithm 1:** SubtreeJump (STJ) proposal algorithm.

0 Excluding the root and its direct descendants, pick a source node $i$ in $\tau$ uniformly at random, i.e., with probability $1/(2n-4)$;
1 Determine the source's parent, $p_i$, and compute $h(p_i)$;
2 Construct the set of destination nodes $\mathbf{D_i} = \{d \in \mathbf{D_i} : h(d) = h(p_i)\}$;
3 For all $k \in \mathbf{D_i}$, compute $\Delta_{ik}$ and $\tilde{\Delta}_{ik} = \frac{\Delta_{ik}}{\max_k \Delta_{ik}}$ respectively;
4 For some fixed $\alpha \in \mathbb{R}$, pick a node $j \in \mathbf{D_i}$ with probability $\Pr(j) = \frac{\tilde{\Delta}_{ij}^{\alpha}}{\sum_{d \in \mathbf{D_i}} \tilde{\Delta}_{id}^{\alpha}}$;
5 Prune the tree at $p_i$ and regraft the resulting subtree at $p_j$, creating a new tree $\tau'$.

We note that STJ is not a symmetric proposal, as for two arbitrary nodes $i$ and $j$ the sets $\mathbf{D_i}$ and $\mathbf{D_j}$ need not coincide. Hence, one needs to compute the *Hastings ratio* $q_\alpha(\tau|\tau')/q_\alpha(\tau'|\tau)$. To perform a reverse move from $\tau'$ back to $\tau$, one needs to pick the

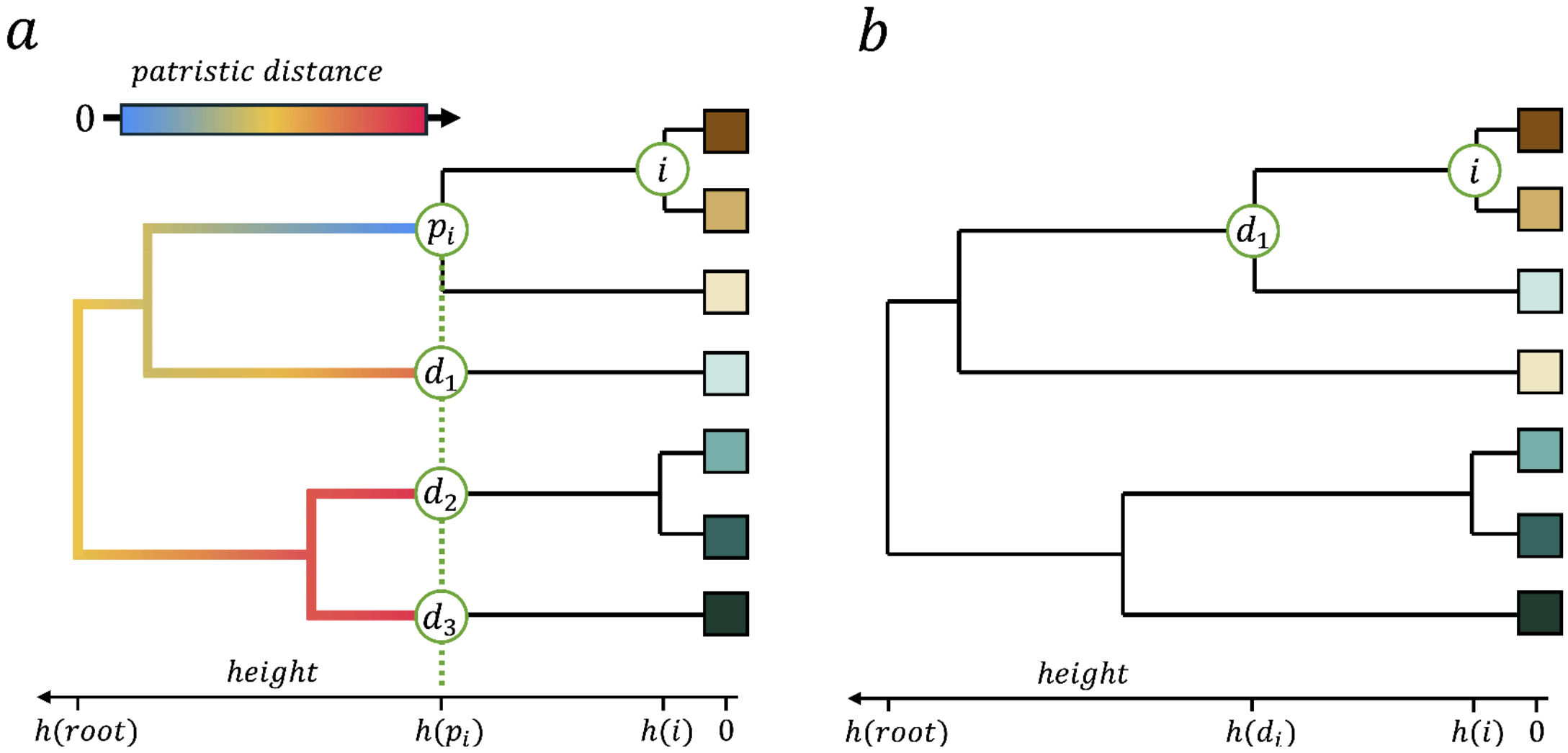

Figure 1: **Schematic representation of SubTreeJump (STJ) a)** Following Algorithm 1, node $i$ is identified for pruning. All nodes at height $h(p_i)$, the height of $i$'s parent node $p_i$, are identified as potential destination nodes, forming the set $D_i$ (here containing $d_1$, $d_2$, and $d_3$). A destination node $j$ is picked at random from $D_i$, with probability $P(i \longrightarrow j)$ proportional to the normalized patristic distance between $p_i$ and $j$ exponentiated to some $\alpha$. Thus, if one chooses $\alpha > 0$, then $P(i \longrightarrow d_1) < P(i \longrightarrow d_2) = P(i \longrightarrow d_3)$, while if $\alpha < 0$, then $P(i \longrightarrow d_1) > P(i \longrightarrow d_2) = P(i \longrightarrow d_3)$. **b)** In this example, $d_1$ is chosen. The subtree defined by $i$ is then pruned and regrafted at $d_1$, conserving $h(i)$ and $h(p_i) = h(d_1)$.

original target node $i$, which is guaranteed to exist in $\mathbf{D_j}$. The Hastings ratio is then

$$\begin{aligned}\frac{q_\alpha(\tau|\tau')}{q_\alpha(\tau'|\tau)} &= \frac{\tilde{\Delta}^\alpha_{ji}}{\sum_{d'\in\mathbf{D_j}} \tilde{\Delta}^\alpha_{jd'}} \Big/ \frac{\tilde{\Delta}^\alpha_{ij}}{\sum_{d\in\mathbf{D_i}} \tilde{\Delta}^\alpha_{id}},\\ &= \frac{\sum_{d\in\mathbf{D_i}} \tilde{\Delta}^\alpha_{id}}{\sum_{d'\in\mathbf{D_j}} \tilde{\Delta}^\alpha_{jd'}} \cdot \frac{\tilde{\Delta}^\alpha_{ji}}{\tilde{\Delta}^\alpha_{ij}}. \end{aligned} \tag{1}$$

We note two limitations of this transition kernel. First, because $\mathbf{H}(\tau)$ is ultimately discrete, not any acceptance probability is attainable. Secondly, STJ on its own does not necessarily induce an irreducible Markov chain on the space of rooted topologies – see Appendix A for proof. This can be easily remedied, however, by combining STJ with transition kernels that operate on the branch lengths.

Like STJ, FNPR was specifically designed to deal with time-calibrated phylogenies, since by proposing new destinations at the same height, it avoids proposing phylogenies that would violate the time constraints inherent to time-calibrated phylogenies. Since FNPR does not change node heights or numbers of lineages, it does not change the density under the coalescent prior, *i.e.* (see Remark 1 in Appendix A). In general, FNPR (like STJ above) requires being paired with transition kernels on the branch lengths in order to generate an ergodic Markov chain and ensure MCMC correctness and efficiency. However, the disconnect between proposals in topological space and branch length space could be undesirable due to it failing to account for the dependence between topology and branch lengths (Yang and Rannala, 2005; Alfaro and Holder, 2006).

### SubTreeLeap

A single adaptive transition kernel to update topology and branch lengths simultaneously would hence be preferable. To this end, we propose *SubTreeLeap* (STL), a transition kernel based on patristic distances. The central idea behind STL is to move a node $i$ to new location in the tree that is at most at (patristic) distance $\delta$ from $i$. To this end, one first draws the distance $\delta$ from a distribution $\kappa(\delta|\sigma)$ indexed by a parameter $\sigma$, henceforth called the *distance kernel*. One then finds the set $\mathbf{D_i}(\delta)$ of all the destination nodes that are at distance $\delta$ from $i$, and picks the destination $j$ uniformly at random from these. We refer to Algorithm 2 and its visualisation in Figure 2 for details.

**Algorithm 2:** SubTreeLeap (STL) transition kernel.

0 Excluding the root, pick a node $i$ in $\tau$ uniformly at random, i.e., with probability $1/(2n-2)$;
1 Draw a patristic distance $\delta$ from the distance kernel $\kappa(\delta|\sigma)$;
2 Find the set of destination nodes $\mathbf{D_i}(\delta) \longleftarrow getDestinations(\tau, i, \delta)$;
3 Pick a node $j \in \mathbf{D_i}(\delta)$ with probability $\Pr(j) = 1/|\mathbf{D_i}(\delta)|$;
4 Prune the tree at $i$ and regraft it at $j$, creating a new tree $\tau'$.
5 **Function** *getDestinations*
  **Input :** A tree $\tau$, a node $i$ and a scalar $\delta$.
  **Output:** A set $\mathbf{D_i}(\delta) = \{d \in \mathbf{D_i}(\delta) : \Delta_{i,d} \leq \delta\}$.
0 Identify the $p_i$ and $s_i$, the parent and sibling nodes of $i$ respectively.
1 From all nodes on the subtree subtended by $s_i$ construct the set $\mathbf{D_i^{(sibling)}} = \{d \in \mathbf{D_i^{(sibling)}} : h(d) = h(p_i) - \delta\}$;
2 From all nodes ancestral to $p_i$ construct the set $\mathbf{D_i^{(ancestral)}} = \{d \in \mathbf{D_i^{(ancestral)}} : h(d) = h(p_i) + \delta\}$;
3 From all other nodes construct the set $\mathbf{D_i^{(other)}} = \{d \in \mathbf{D_i^{(other)}} : h(d) = 2 * h(mrca_{p_i,d}) - (h(p_i) + \delta)\}$;
4 Construct the union set $\mathbf{D_i^{(union)}} = \mathbf{D_i^{(sibling)}} \cup \mathbf{D_i^{(ancestral)}} \cup \mathbf{D_i^{(other)}}$ ;
5 Construct the set of destination nodes which would violate temporal constraints $\mathbf{D_i^{(invalid)}} = \{d \in \mathbf{D_i^{(invalid)}} : h(d) \in \mathbf{D_i^{(union)}} \wedge h(d) < h(i)\}$;
6 **return** $\mathbf{D_i}(\delta) = \mathbf{D_i^{(union)}} \setminus \mathbf{D_i^{(invalid)}}$

STL is not symmetric, hence we need to compute the Hastings ratio $q_\sigma(\tau|\tau')/q_\sigma(\tau'|\tau)$. In order to move back to $\tau$ from $\tau'$, one would first need to draw the same distance $\delta$ from $\kappa(\cdot)$. The original node $i$ is guaranteed to exist in the set of destinations $\mathbf{D_j}$, but $\mathbf{D_j} \neq \mathbf{D_i}$. The

Hastings ratio equals

$$\frac{q_\sigma(\tau|\tau')}{q_\sigma(\tau'|\tau)} = \frac{1}{|\mathbf{D_j}(\delta')|} \Big/ \frac{1}{|\mathbf{D_i}(\delta)|}, \\ = \frac{|\mathbf{D_i}(\delta)|}{|\mathbf{D_j}(\delta')|} \tag{2}$$

Note that one needs to draw the exact same distance $\delta' = \delta$ to be able to get the original node in the destination set and hence produce $\tau$ from $\tau'$. This means the densities $\kappa(\delta'|\sigma)$ and $\kappa(\delta|\sigma)$ cancel out in the Hastings ratio, leaving only the discrete component to be computed. This is independent of the choice of distance kernel, as long as $\kappa(\cdot|\sigma)$ is strictly positive and unbounded (see Remark 2 in Appendix A). STL is ergodic on $\mathbf{\Psi}$, which establishes its suitability for use as the sole phylogenetic transition kernel in an MCMC analysis. We provide a proof in Appendix A (Theorem 1).

This construction is complicated and deserves additional consideration. Define $g_i$ as the grandparent node of $i$. Depending on the magnitude of $\delta$, a range of rearrangements is possible:

1. *Slide move*: If $h(p_i) - \delta > h(s_i)$, or if $h(p_i) + \delta < h(g_i)$, then the destination node $j$ of $i$ could end up on the branch subtending or subtended by $p_i$. In this case, no topological rearrangement occurred, and STL effectively rescaled branch lengths. The Hastings Ratio for a Slide move is 1.

2. *Same subtree move*: If $h(p_i) - \delta < s_i$, then $j$ could end up on the subtree defined by $s_i$. In this case, only the subtree of $p_i$ was rearranged, leaving the rest of the phylogeny unchanged.

3. *Cross-subtree move*: If $h(p_i) - \delta > g_i$, then $j$ could end up anywhere else in the tree not subtended by $p_i$.

4. *Root-altering move*: If $\delta > \max_j \Delta_{p_i,j}$, there will be no destinations on the opposite subtree. In this case, there is only one destination node, "above the root", and $p_i$ becomes the root. This is in stark contrast with the default set of transition kernels available in BEAST X (Baele et al., 2025), where one typically employs a specific transition kernel to propose changes to the height of the root. As a trade-off however, STL will only change the root height occasionally, with the probability of such a move decreasing with the number of taxa.

Note that all cases mentioned above are subject to temporal constraints, i.e. a destination node $j$ must be higher on the phylogeny than $i$: $h(j) < h(i)$.

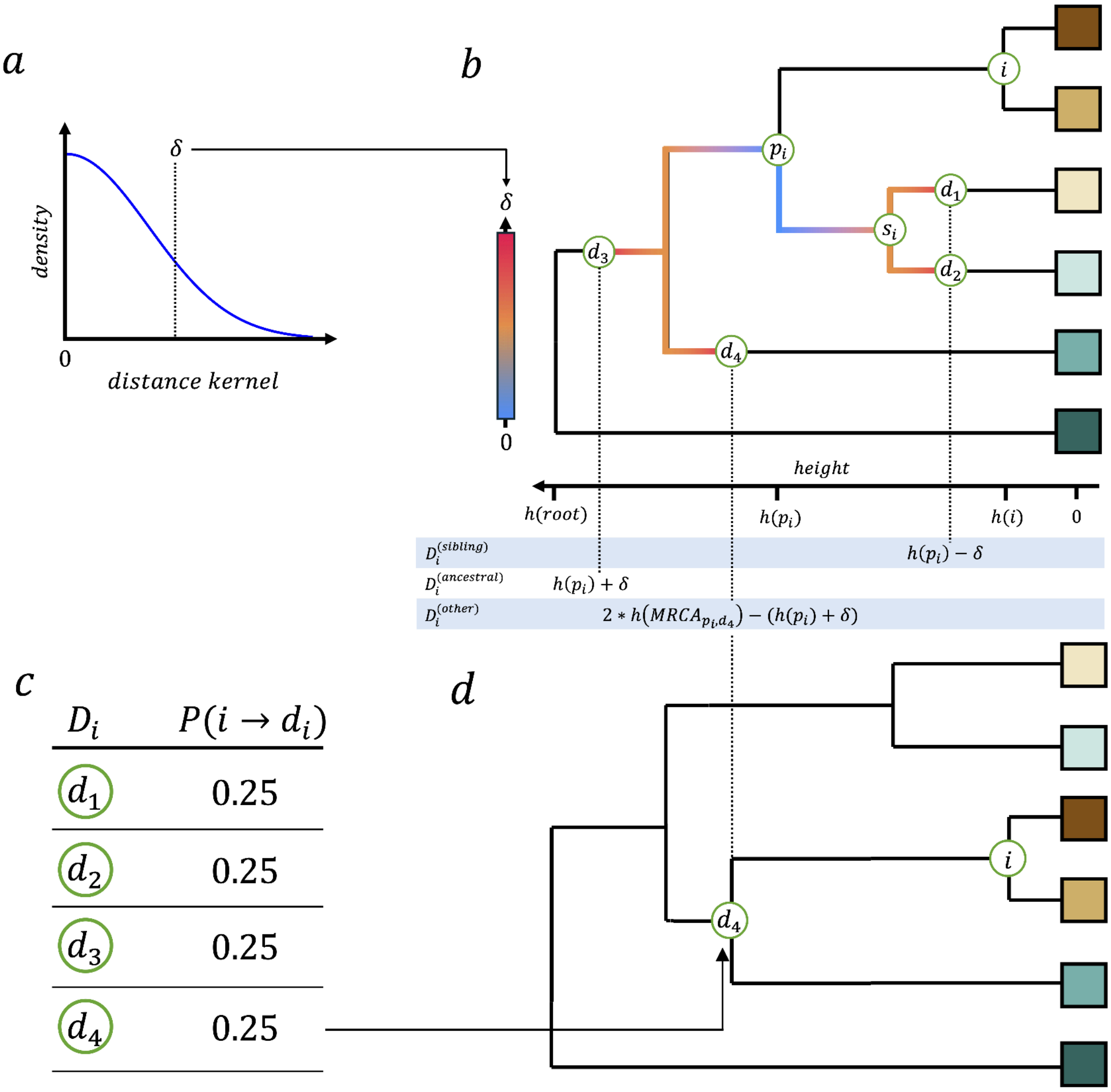

Figure 2: **Schematic representation of SubTreeLeap (Alg. 2) a)** A patristic distance $\delta$ is drawn from a strictly positive distance kernel. **b)** Node $i$ is selected at random for pruning. All nodes at patristic distance $\delta$ from the parent of $i$, $p_i$, are identified as potential destination nodes, forming the different subsets of the destination set $D_i$ (see Algorithm 2). **c)** A node is picked at random from $D_i$ with equal probability (in this case: 0.25). **d)** The subtree defined by $i$ is thus pruned and regrafted to the destination node, while conserving its height $h(i)$.

## Convergence assessment in topological MCMC

A crucial step in obtaining expectations through MCMC is convergence assessment, i.e. ascertaining whether the algorithm has failed to converge to the desired target distribution (Cowles and Carlin, 1996). For regular parameter spaces – such as Euclidean spaces – convergence is often assessed visually, by monitoring the evolution of the sampled parameter values over time, or by the use of convergence statistics such as the Gelman-Rubin statistic (Gelman and Rubin, 1992) or the Geweke diagnostic (Geweke, 1992). Phylogenetics however poses a particularly hard challenge due to the non-standard form of the parameter space (St. John, 2017). Visual methods can be challenging because directly visualising the evolution of entire topologies for thousands of samples is not feasible. Convergence statistics meanwhile typically rely on comparing variability between chains (e.g. the Gelman-Rubin metric) or sections of chains (e.g. the Geweke diagnostic), which cannot be straightforwardly generalised to the case of phylogenetic trees without making arguably arbitrary decisions on how such variability should be quantified. Nonetheless, there exists a substantial body of research that aims to create novel convergence diagnostics or extend existing ones for the case of MCMC in phylogenetics.

### Visualising (convergence in) tree space

Lanfear et al. (2016) introduced the topological trace plot, an extension of the "trace graph" often utilised in more classical MCMC diagnostic settings, where the value of the sampled parameter is plotted against the MCMC iteration. The value being graphed by Lanfear et al. (2016) is the phylogenetic distance (PD) from a reference tree to the tree sampled at each iteration. This type of visualisation requires two somewhat arbitrary choices to be made: which PD metric to use and which reference tree to choose. A range of PD metics have been developed, e.g. the Robinson-Foulds distance (Robinson and Foulds, 1981), the Subtree-Prune-Regraft distance (Beiko and Hamilton, 2006), or the quartet distance (Brodal et al., 2001). Practical choices for the reference tree include a random posterior tree sample in the MCMC analysis and a consensus tree of the posterior tree samples collected (Lanfear et al., 2016). Both of these decisions can have a substantial impact on conclusions (Brusselmans et al., 2024), and there are no established best practices at the moment.

An alternative visual method that resolves the reference tree issue involves computing all pairwise distances between sampled trees, performing a dimensionality reduction technique such as principal coordinates analysis – also known as (classic) multidimensional scaling (MDS) – on the produced distance matrix, and plotting the resulting coordinates in 2D space (Hillis et al., 2005). This approach is capable of picking up on multimodal posteriors in tree space and between-chain convergence issues (Smith, 2022; Gao et al., 2026), but is considerably more computationally expensive, as the number of PDs to compute grows quadratically with sample size as opposed to linearly for the trace plot.

### Convergence statistics in tree space

Lakner et al. (2008) introduced the average standard deviation of split frequencies (ASDSF), which is calculated by comparing split or clade frequencies across multiple independent MCMC replicate analyses started from different randomly chosen starting trees. Specifically, let $f_i^j$ be the fraction of the trees from run $i$ containing the split or clade (taxon bipartition) $j$, then the ASDSF is defined as

$$\text{ASDSF} = \frac{1}{M} \sum_j \left( \sqrt{\frac{1}{N-1} \sum_i (f_i^j - \bar{f}^j)^2}, \right) \tag{3}$$

where $N$ is the number of runs, $M$ is the total number of splits or clades present in the tree samples being compared, and $\bar{f}^j$ is the mean fraction of trees across all runs containing split or clade $j$. The ASDSF is expected to decrease as the different chains converge to the same stationary distribution. Lakner et al. (2008) report that they were able to obtain reasonably accurate estimates of the posterior distribution in the reference runs (ASDSF $\leq$ 0.02) for 6 out of their 20 data sets, and that ASDSF values reached below 0.05 for 3 more, while another three had ASDSF values below 0.10 (the remainder were even more heterogeneous at 20 million generations). It is thus important to emphasize that no single cut-off for the ASDSF should be employed in practice, since the value at which it stabilises as the chain progresses will depend on the dataset. For evaluating transition kernels, it is therefore more relevant to study the shape of the ASDSF curves, rather than their absolute values. Alternative convergence diagnostics exist, such as the topological Gelman-Rubin statistic (Whidden and Matsen, 2015), which substitutes the between- and within-chain variances needed to compute the classical version with the between- and within-chain deviations of pairwise phylogenetic distances. This, in turn, makes the value of the topological Gelman-Rubin statistic dependent on the choice of PD metric, with all the previously mentioned considerations.

### Golden MCMC analyses to characterise tree space

While the previously described methods can be used to evaluate whether the sampling distribution has reached stationarity, it is hard to tell whether convergence to the target posterior has been reached. If a chain reaches pseudo-convergence, whereby it attains stationarity without having truly converged (such as when it gets stuck in a local maximum), it can very well pass all "convergence checks" (Geyer, 2011). When dealing with real-world data, one does not know the true posterior and thus no definitive reference against which to compare exists. To this end, the use of "golden MCMC analyses" or "reference samples", which are obtained through careful and detailed exploration of a given data set, has been advocated in phylogenetic inference. The purpose of these analyses is to allow several MCMC chains to explore tree space so thoroughly that undetected pseudo-convergence becomes improbable, at least within the realm of what our current methods are capable of. One could summarise the use of golden runs to characterise tree space as follows: firstly, perform several runs under different starting conditions. Secondly, ensure all runs reach stationarity at the same distribution. Finally, run the chains for a sufficiently long time to ensure thorough sampling of the posterior.

Lakner et al. (2008) used at least six independent / parallel Metropolis-coupled MCMC (i.e., $MC^3$) analyses from MrBayes (Ronquist et al., 2012), starting from different random starting phylogenies and each of them using four Metropolis-coupled chains under the default MrBayes heating scheme and run until the standard deviation of split frequencies was 0.005. Höhna et al. (2008) also note that in practice the true target distribution of clades is unknown, and that in such cases a set of extensive / extremely long "golden" MCMC analyses are performed to estimate this distribution as accurately as possible. Höhna and Drummond (2012) relied on the similarity measurement between the sampled distribution and the target distribution that was introduced by Lakner et al. (2008) and Höhna et al. (2008), and performed 10 golden MCMC analyses of 1 billion iterations for each of their data sets with the BEAST 1.7 (Drummond et al., 2012) default transition kernels and priors. Employing an approach that holds the middle between those of Lakner et al. (2008) and Höhna et al. (2008), Whidden and Matsen (2015) computed large "golden run" posterior samples for each of their data sets by repeatedly running single-chain MCMC analyses for 1 billion iterations. Following Höhna and Drummond (2012) and Whidden and Matsen (2015), studies like Whidden et al. (2019) and Gao et al. (2026) performed 10 single-chain MCMC analysis replicates which were run for 1 billion iterations but sampled much more often than is commonly done, i.e. every 100 iterations, and assessed that the estimated split frequency error was below 0.015% for each of their data sets, suggesting that the various golden runs sampled the same split frequencies.

### Post-convergence tree sampling efficiency

Once no convergence problems could be detected, we can consider the separate issue of statistical efficiency (i.e., mixing) in tree space. This is fundamentally related to the number of effective (independent) samples from the target distribution obtained. A higher effective sample size (ESS) implies a lower autocorrelation time, thus suggesting a higher rate at which information on the posterior distribution is gathered through the sampling process. Defining autocorrelation for phylogenetic trees is not straightforward, however. Lanfear et al. (2016) argued that autocorrelation time can be computed from pairwise distances between sampled trees, based on the analogy between expected squared pairwise distances between sampled trees and the covariance of samples of continuous variables. Magee et al. (2024) introduced a number of alternative topological ESS estimators and performed extensive simulations to assess their performance, finding the best performing estimators to be the "split frequency ESS", "Fréchet correlation ESS", and "pseudo-ESS". We refer to Magee et al. (2024) for an explanation on the workings of the latter two. These estimators, with the exception of the split frequency ESS, also rely on an arbitrary choice of phylogenetic distance metric, the effect of which on their estimates is large (Brusselmans et al., 2024), but has not systematically been characterized. One can obtain the split-frequency ESS by computing the sample variance and using an estimator of the asymptotic variance of the Markov chain (Magee et al., 2024):

$$\widehat{\text{ESS}} = n \frac{\hat{\sigma}_P^2}{\hat{\sigma}_L^2}, \tag{4}$$

where the $n$ is the number of samples, and $\hat{\sigma}_L^2$ and $\hat{\sigma}_P^2$ are estimates of the asymptotic variance and posterior variance of the sample mean respectively. The posterior variance is computed by reducing the trees to binary vectors indicating the presence or absence of possible splits, and applying a Fréchet generalization of variance based on the Euclidean distance:

$$\hat{\sigma}_P^2 = \frac{1}{n-1} \sum_{i=1}^{n} ||\mathbf{X}_i - \bar{\mathbf{X}}||_k^2, \tag{5}$$

where $\mathbf{X}_i$ is the binary vector of splits in a given tree, $\bar{\mathbf{X}}$ is the vector of arithmetic mean frequencies of splits over the entire sample, and $k$ is the total number of observed splits. The asymptotic variance is estimated using the batch means estimator, computed as follows, starting from:

$$\gamma_B = \frac{B}{a-1} \sum_{i=1}^{a} ||\mathbf{Y}_i - \bar{\mathbf{X}}||_k^2 \tag{6}$$

where $B$ is the batch size, $a = \lfloor \frac{n}{B} \rfloor$ is the number of batches, and $\mathbf{Y}_{i=1,\dots,a}$ are the vectors of mean split frequencies in a batch. A naive estimator of the asymptotic variance uses batch size $\sqrt{n}$, so $\hat{\sigma}_L^2 = \gamma_{\sqrt{n}}$, but Vats and Knudson (2021) use a replicated lugsail correction which estimates the asymptotic variance as $\hat{\sigma}_L^2 = 2\gamma_{\sqrt{n}} - \gamma_{\frac{\sqrt{n}}{3}}$, as this has been shown through simulation to provide better estimates when different sections of the sample explore different modes of the distribution (Argon and Andradóttir, 2006).

### Experimental setup

We compare the performance of two tree transition kernel mixes, which we will refer to as the "classic mix" and the "STL mix". The classic mix uses a number of different transition kernels (shown in Table 1), while the STL mix only uses two: STL and STJ. Transition kernel weights are fixed to default BEAUti X (v10.5.0) settings, thus with fixed values for the kernels in the classic mix, and scaling linearly with data set size for kernels in the STL mix. In order to minimise model noise from other transition kernels and allow for a more direct comparison of the two transition kernel mixes, we fixed the parameter values in all substitution models to their estimates from preliminary analyses.

| Kernel | Weight | Role |
|---|---|---|
| **Classic mix** | | |
| SubTreeSlide | 30 | Slides a subtree uniformly along a branch |
| NarrowExchange | 30 | Swaps a subtree with its uncle |
| WideExchange | 3 | Swaps two random subtrees |
| WilsonBalding | 3 | Prunes a subtree and attaches it to a new branch |
| Uniform | 30 | Changes one internal node height |
| Scale | 3 | Changes the root node height |
| **STL mix** | | |
| SubTreeJump | $0.1 \times$ #taxa | STJ operation (Algorithm 1) |
| SubTreeLeap | #taxa | STL operation (Algorithm 2) |

Table 1: STL mix is the default tree transition kernel composition in BEAST X (Baele et al., 2025), with the classic mix having been the default in earlier versions of the BEAST software package (Drummond and Rambaut, 2007; Drummond et al., 2012; Suchard et al., 2018).

For each data set, we perform 4 golden MCMC analyses of 1 billion iterations each, using both classic and STL kernels, thus totalling 8 MCMC analyses per data set. We performed these analyses using BEAST X (Baele et al., 2025), supported by the BEAGLE 4 (Gangavarapu et al., 2026) high-performance library for statistical phylogenetics. We assessed convergence and mixing for continuous model parameters and statistics using Tracer v1.7.2 (Rambaut et al., 2018).

We compare the convergence rate of the classic and STL kernel mixes by graphing the ASDSF over time for each transition kernel mix. To visualise explored tree space post-convergence, we compute pairwise Robinson-Foulds (Robinson and Foulds, 1981) distances between 300 equally spaced sampled trees from the last 500 million iterations (latter half of the chain) of each of the 8 runs. We perform classic MDS and graph the first six dimensions (Smith, 2022). Additionally, we repeat the visualisation using the approximate SPR distance and highlight whether and how the choice of PD metric influences our conclusions. Evaluation of post-convergence sampling efficiency was performed using the split-frequency ESS, graphing its value over time starting at the 500 millionth iteration. This implies we consider the first half of the chain as burn-in. While this is likely a stringent cut-off, the sampling distribution should be unchanging after reaching stationarity, and thus not impact estimates of ESS increase per iteration. Leveraging the use of extremely long golden analyses by evaluating sampling efficiency well after the chain has reached stationarity thus avoids mistakenly including burn-in in those analyses. Additional dataset-specific analyses that were performed to investigate further peculiarities are described in their respective sections. All analyses of the BEAST X output were performed in R v4.3.0 (R Core Team, 2025) using the packages **ape** (Paradis and Schliep, 2019), **phangorn** (Schliep, 2011), and **treess** (Magee et al., 2024).

## Data and models

In order to cover a meaningful range of diversity in terms of data set complexity and tree shape, we utilised three substantially different data sets with increasingly larger numbers of both sequences and sites. A summary of the computationally relevant information for each data set can be found in Table 2.

The data set of Worobey et al. (2008) (henceforth referred to as "HIV") compiles 162 HIV type 1 sequences from 12 concatenated gene fragments. This is our smallest data set, characterised by a low number of sites (994) and collection dates that span 46 years. Grubaugh et al. (2019)'s 283 full-genome Zika virus data set ("ZIKV") functions as our mid-sized example. These full genomes are substantially longer than the HIV sequence fragments with 10 269 sites and cover a much shorter timespan (5 years). Our largest data set is from Feinauer et al. (2024), a 449 taxon set of mitochondrial genomes from Scandinavian brown bears ("BEAR"), augmented with conserved permafrost samples with dates estimated to be from up to 100 thousand years ago. This makes the trees estimated from these data effectively ultrametric, with the exception of those few ancient sequences. As such, we cover an adequate range of phylogenetic behaviours and tree topologies.

| Data set | # sequences | # sites | Time span (years) |
|---|---|---|---|
| HIV | 162 | 994 | 1959-2005 (46) |
| ZIKV | 283 | 10 269 | 2013-2018 (5) |
| BEAR | 449 | 15 118 | ± 100 000BC - 2024 (± 100 000) |

Table 2: **Data set specifications.** This table reports the size, length, and temporal distribution of the tree case study datasets utilized in the analyses. The datasets are ordered in increasing complexity.

The same phylogenetic models were used for all three data sets. We employed an HKY+Γ4 (Hasegawa et al., 1985; Yang, 1994) substitution model with an uncorrelated relaxed molecular clock with an underlying lognormal distribution (Drummond et al., 2006) and a non-parametric skygrid coalescent tree prior (Gill et al., 2013). We used a Hamiltonian Monte Carlo (HMC) transition kernel to efficiently estimate the parameters of the skygrid (Baele et al., 2020).

# Results

## HIV

Employing the STL kernel mix in the HIV analyses substantially increased topological convergence rate between replicate analyses compared to those utilising the classic kernel mix (Figure 3A): the ASDSF of the former stabilises around 200 million iterations, while the latter takes approximately 400 million iterations to stabilise, and does so at a higher ASDSF value that – contrary to the STL kernel mix — shows no consistently downward trend. Topological mixing is better for the STL kernel mix as well, as the split-frequency ESS increases remarkably faster compared to the classic kernel mix (Figure 3B). A notable oddity is classic run #2, whose ESS increase matches STL runs up until around 300 million iterations post burn-in, after which a sharp decrease is observed as it joins the ESS accumulation curves of the other classic kernel mix runs.

Despite the faster convergence and better mixing for the STL kernel mix, it produces a practically indistinguishable exploration of posterior tree space to that obtained with the classic kernel mix based on an RF distance-based MDS projection (Figure 3C). Generally speaking, the explored posterior tree space is homogeneous for the HIV data set, and there is no noticeably consistent difference between the regions explored by both kernel mixes, or regions of posterior tree space not explored at all by any of the replicate analyses. We refer to Supplementary Figure S1 for additional RF distance-based visualisations per kernel mix type, and to Supplementary Figures S2 and S3 for highly similar results using SPR distance-based visualisations that illustrate that our findings do not depend on the choice of phylogenetic distance metric.

However, the manner in which posterior tree space is explored under both kernel mixes is clearly different, as can be seen in Figure 4. The STL kernel mix leads to a more thorough posterior tree space exploration compared to the classic kernel mix, and does so in a consistent manner, i.e. each subsequent set of 100 million iterations explores a larger part of tree space compared to the classic kernel mix. The classic kernel mix moves much slower through this space and as a result consistently remains more localized (in posterior tree space). While the final sampled posterior tree space is nearly identical under both kernel mixes, it will hence take analyses with the classic kernel mix substantially longer to fully explore the space compared to the STL kernel mix.

We can use the observations in Figure 4 to further explain the counter-intuitive result for classic run #2 seen in Figure 3B, which implies that at some point in the MCMC

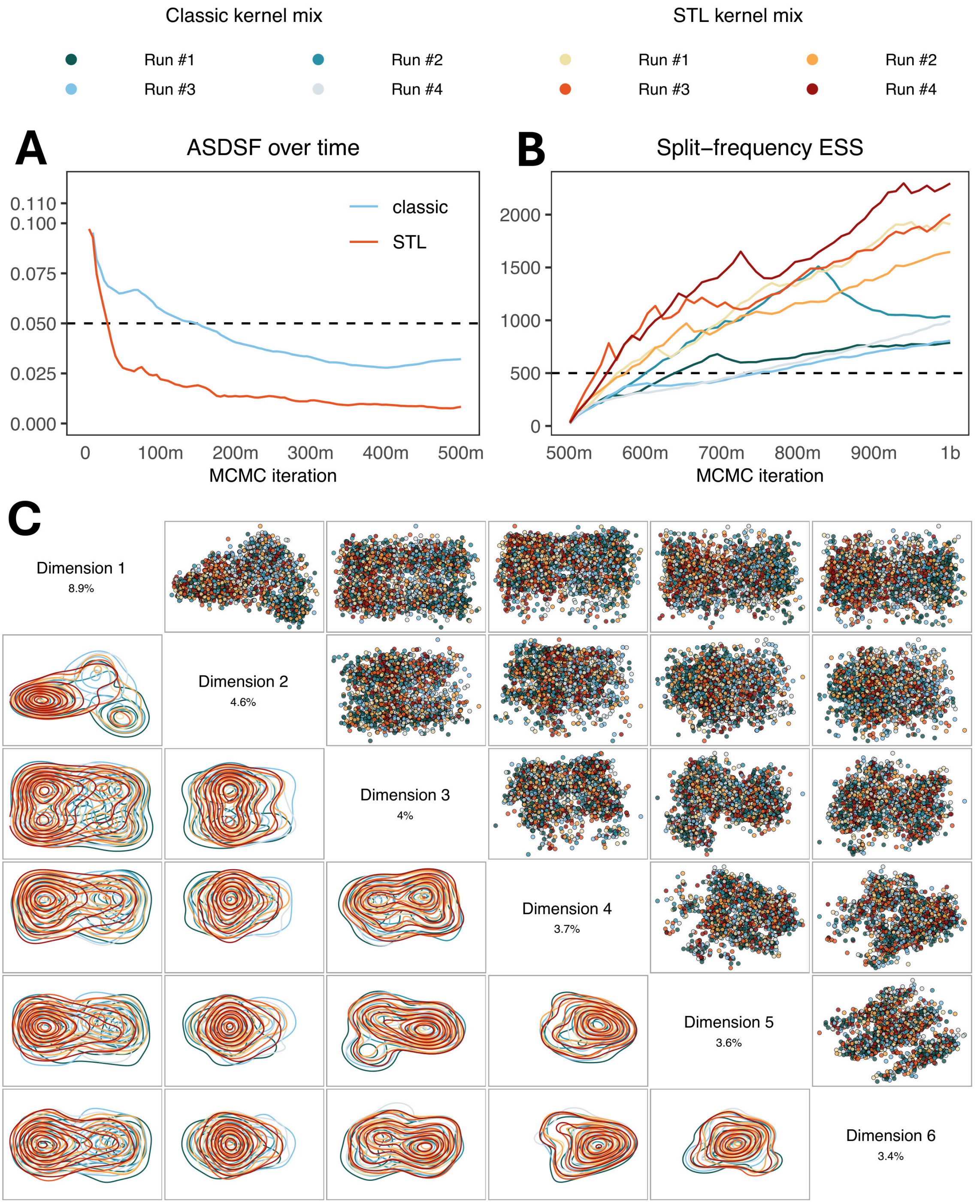

Figure 3: **Topological convergence and mixing analysis for the HIV data set. A)** Average standard deviation of split frequencies (ASDSF) for four replicate analyses of both classic and STL kernel mixes shows an increased topological convergence rate for the STL kernel mix. **B)** The multidimensional effective sample size (ESS) after burn-in based on split frequency vectors increases at a higher rate for replicate analyses with the STL kernel mix compared to the classic kernel mix. **C)** A 6-dimensional MDS performed on RF distances between 300 equally spaced posterior samples of each run (after burn-in) shows a largely homogenous explored tree space by both kernel types. Above the diagonal are trees MDS coordinates for each tree, below the diagonal are 2D kernel density contour lines of said coordinates. Percentages shown are the proportion of explained variability by each dimension.

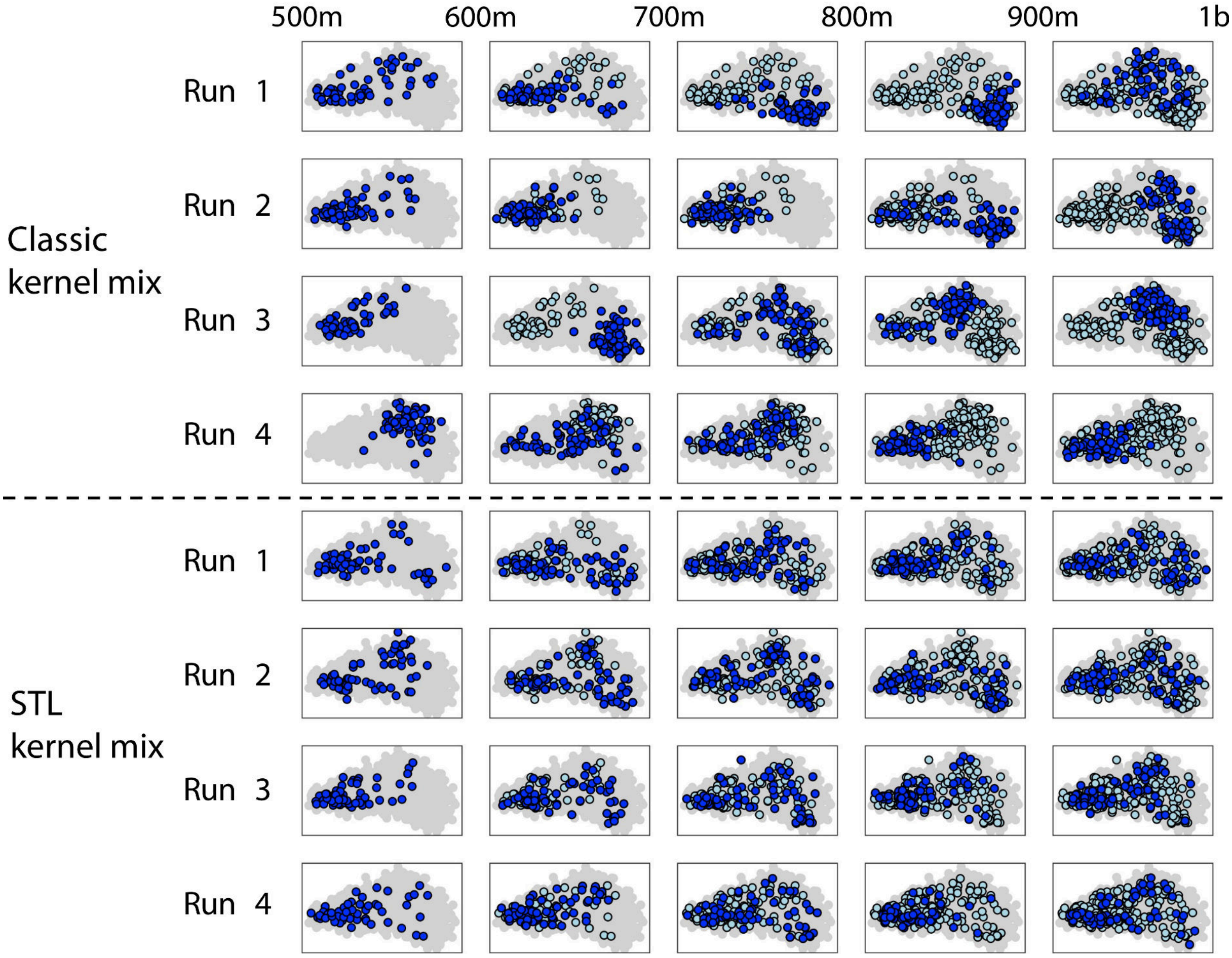

Figure 4: **Dimensions 1 and 2 of RF distance-based PCoA posterior tree space for the HIV data set.** After discarding the first 500 million iterations as burn-in, this figure shows how both classic and STL kernel mixes explore posterior tree space throughout the following 500 million iterations. Dark blue points show posterior trees sampled during a given window of 100 million iterations. Light blue dots show all trees sampled up to that point. Grey background dots show the entire posterior tree space for the post-burn-in 500 million iterations for all runs combined. Runs that employ the classic kernel mix show a clearly temporally heterogeneous posterior tree space exploration: within any slice of 100 million iterations, the chain only explores part of the tree space, often remaining "stuck" in a particular area for a prolonged segment (e.g. classic run #2.). In contrast, runs using the STL kernel have noticeably better mixing: in each slice of 100 million iterations, the chain covers the entirety of tree space much better. This is consistent with the higher split-frequency ESS's observed for STL kernels in Figure 3B, as a given number of posterior samples from the STL kernel runs contain more information than an equal number of samples from classic kernel mix runs.

analysis – after 500 million iterations were discarded as burn-in, and hence after 800 million iterations in the posterior tree space exploration – a different (previously unvisited) region in posterior tree space is being explored. We can see this in Supplementary Figures S4 and S5; the timing of the ESS decline observed in Figure 3B (around 300 million iterations post-burn-in) follows the first move of the MCMC analysis to the right side of the posterior tree space in the MDS projection. This pattern is also observed in the other analyses that show (lesser) ESS declines, e.g. STL runs #3 and #4. These slumps in topological ESS values thus reflect the exploration of new areas or modes of posterior tree space, and the Markov chain that subsequently remains localized in that region of tree space for an extensive number of iterations until it “escapes” said region to continue exploring tree space in a more homogeneous manner again.

**Zika virus (ZIKV)**

The results of the ZIKV analyses show similar patterns as those observed in the HIV analyses, although the differences between the classic and STL kernel mixes are less pronounced. The STL kernel mix again produces an ASDSF that stabilises at a lower value than the classic kernel mix, and the split-frequency ESS again increases at a higher rate compared to the classic kernel mix (Figure 5A,B). However, tree space is notably multimodal, with two to six distinctly visible modes depending on which MDS dimensions are being projected (Figure 5C and Supplementary Figure S6). None of the (classic and STL) replicate analyses have difficulties jumping between the modes. As such, each analysis replicate visits each mode regularly, indicating that whatever differences in topology characterize each mode are easily bridged by both the classic and STL transition kernel mixes.

To further investigate the multimodality in RF posterior tree space, we also assess this space through an SPR lens in Supplementary Figures S7 and S8. Our previous work suggests that RF distances can pick up on phylogenetically meaningful differences between trees (Brusselmans et al., 2024; Gao et al., 2026), even though, as Smith (2022) indicated, they do occasionally throw out “false flags” as only a handful of unstable sequences can give the impression of an unstable phylogeny when viewed through RF space (Gao et al., 2026). For this ZIKV example, Supplementary Figures S7 and S8 show a homogeneous posterior tree space when using the SPR phylogenetic distance for the MDS projection, indicating that both kernel mixes explore tree space equally well but also toning down the “severity” of the multimodality observed in RF tree space, meaning that the RF modes represent small changes in topology that are easily traversed by a single transition kernel application.

This is confirmed through a visualisation of how the different kernel mixes keep on exploring posterior tree space, which is highly similar for both mixes (Figure 6) in that each subsequent set of 100 million iterations explores all areas of tree space. Further, both classic and STL kernel mixes are seen to move between the multiple modes in RF space.

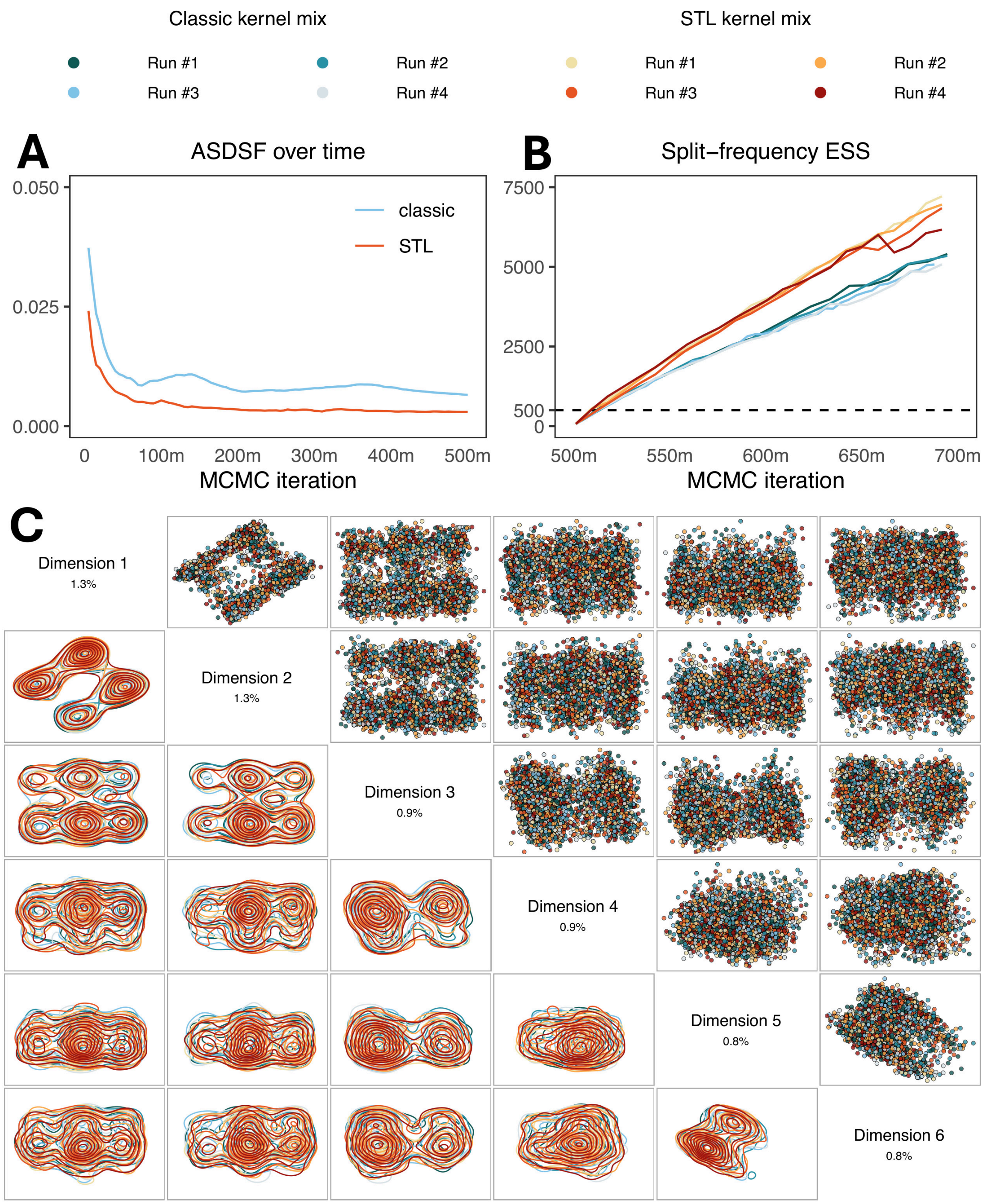

Figure 5: **Topological convergence and mixing analysis for the ZIKV data set. A)** Average standard deviation of split frequencies for four runs of both classic and STL kernels show similar but slightly increased topological convergence rate for the latter. **B)** The multidimensional effective sample size (after burn-in) based on split frequency vectors increases at a higher rate for runs with STL kernels compared to the classic mix. Extended version can be found in Figure S14A. **C)** A 6-dimensional MDS performed on RF-distances between 300 equally spaced posterior samples of each run (after burn-in) show a bimodal tree space explored by both kernel types. Above the diagonal are trees MDS coordinates for each tree, below the diagonal are 2D kernel density contour lines of said coordinates. Percentages shown are the proportion of explained variability by each dimension.

While the final sampled posterior tree space is nearly identical under both kernel mixes, we can thus again conclude that the STL kernel mix produces more consistent posterior tree space exploration between replicate analyses (Figure 5A) and more rapidly gathers a set of independent samples from tree space (Figure 5B).

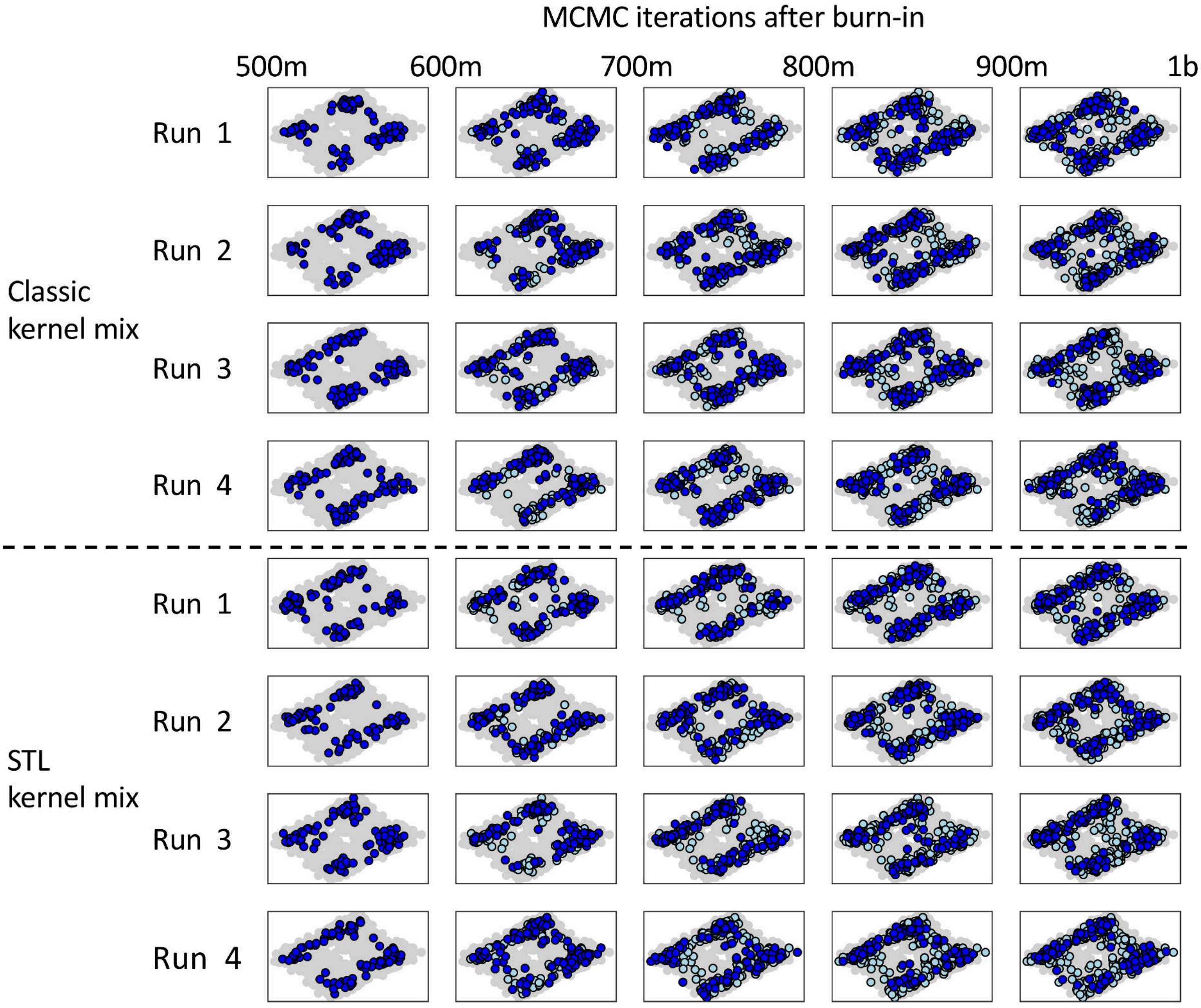

Figure 6: **Dimensions 1 and 2 of RF distance-based PCoA posterior tree space for the ZIKV data set.** After discarding the first 500 million iterations as burn-in, this figure shows how both classic and STL kernel mixes explore posterior tree space throughout the following 500 million iterations. Dark blue points show posterior trees sampled during a given window of 100 million iterations. Light blue dots show all trees sampled up to that point. Grey background dots show the entire posterior tree space for the post-burn-in 500 million iterations for all runs combined. Both kernel mixes show temporally consistent homogeneous explorations of posterior tree space and equally frequent transitions between the modes in the RF distance-based MDS projection.

### Brown bears (BEAR)

For the BEAR dataset, the ASDSF decreases faster and to a lower level for the classic kernel mix compared to the STL kernel mix during the first 100 million iterations, indicating faster convergence between analysis replicates (Figure 7A). In terms of sampling efficiency, the split-frequency ESS increases at a higher rate using the STL kernel mix compared to using the classic kernel mix (Figure 7B). Posterior tree space appears homogeneous, with visible multimodality in the first two projected MDS dimensions (Figure 7C). Again, we observe no difficulties for either kernel mixes in crossing these modes in RF space. As with the ZIKV analysis, the multimodality does not appear in the SPR distance-based MDS projection, as can be seen in Supplementary Figures S7 and S8.

A visualisation of how the different kernel mixes keep on exploring posterior tree space after the burn-in shows a highly similar scenario for both kernel mixes (Supplementary Figure S12) in that each subsequent set of 100 million iterations explores all areas of tree space. Further, both classic and STL kernel mixes are seen to move between the multiple modes in the first two projected MDS dimensions in RF space. However, a closer inspection on the burn-in of the replicate analyses (Figure 8) shows how replicate STL kernel mix analyses remain in a visible burn-in section of tree space for longer than the classic kernel mix analyses, confirming that convergence does indeed take place at a later point for the STL kernel mix on the BEAR data set as shown previously in Figure 7A. To clarify why this occurs, Figure S13 shows how the root age of the BEAR trees reaches stationarity later when using the STL kernel mix compared to the classic kernel mix. Given the vastly different time scale for this data set compared to the HIV and ZIKV analyses, the lack of a dedicated transition kernel to update the root height in the STL kernel mix may prove to be a disadvantage. Given that we aimed at maintaining the same STL kernel mix composition as in the previous examples, we did not consider modifying the STL kernel mix with a specific – and computationally cheap – transition kernel to rectify this, although it would be straightforward to do so.

# Discussion

We have shown that the STL kernel mix consistently explores the same posterior space as the classic kernel mix, as seen in the MDS projections for each dataset. There are however limitations to the use of MDS as a posterior tree space visualisation approach. The proportion of total variance accounted for by the first six coordinates of the MDS is consistently low, ranging from 28.2% to 4.5% in the HIV and BEAR data sets respectively. This suggests there is a large degree of topological variation that is not accounted for in the visualisations used here. Despite this, we can have some degree of confidence in the representativeness of the MDS when we cross-reference it with other supporting analyses and visualisations. As mentioned before, the trends observed in the ASDSF and split-frequency ESS values have parallels in the MDS visualisations, as seen in the analogy between Figure 4 and Figure 3B, where slower ESS increases were reflected in slower tree-space traversal, and decreases in ESS were reflected in the "discovery" of new tree-space modes. We can bypass the need for dimensionality reduction by visualising the pairwise distances directly, as in Supplementary Figures S4 and S5. Supplementary Figure S5 shows how the pairwise phylogenetic distances between replicate analyses using the same kernel mixes are indistinguishable from distances between analyses using different kernels, supporting the homogeneous tree space seen in Figure 3C.

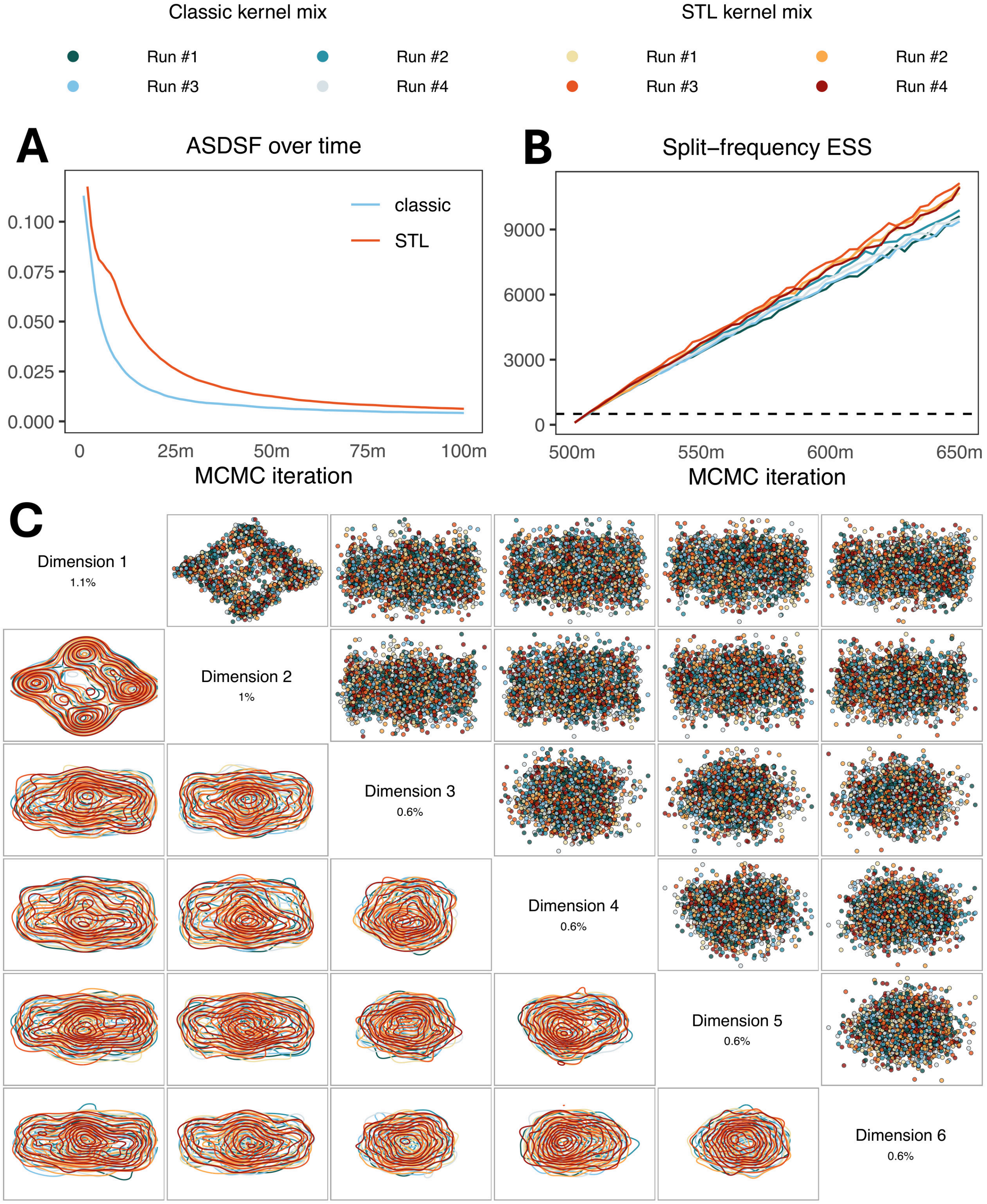

Figure 7: **Topological convergence and mixing analysis for the BEAR data set. A)** Average standard deviation of split frequencies for four runs of both classic and STL kernels show similar but slightly decreased topological convergence rate for the latter. **B)** The multidimensional effective sample size (after burn-in) based on split frequency vectors increases at a slightly higher rate for runs with STL kernels compared to the classic mix. Extended version can be found in Figure S14B. **C)** A 6-dimensional MDS performed on RF-distances between 300 equally spaced posterior samples of each run (after burn-in) show a largely homogeneous tree space explored by both kernel types. Above the diagonal are trees MDS coordinates for each tree, below the diagonal are 2D kernel density contour lines of the coordinates. Percentages shown are the proportion of explained variability by each dimension.

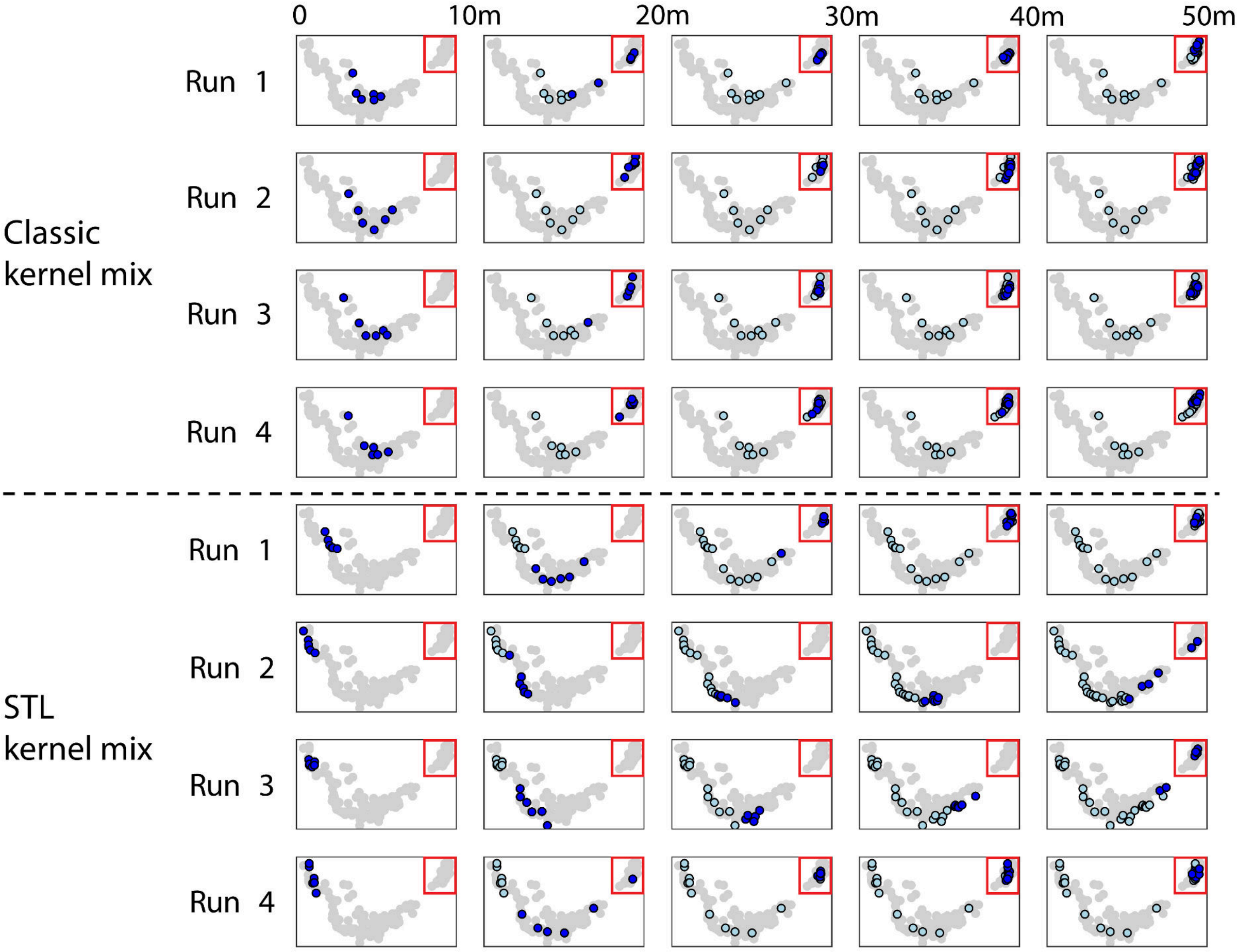

Figure 8: **Dimensions 1 and 2 of RF distance-based PCoA burn-in tree space for the BEAR data set.** Without discarding any of the iterations as burn-in, this figure shows how both classic and STL kernel mixes make their way through tree space during the first 50 million iterations of their 4 replicate analyses. Classic kernel analyses are seen to consistently converge towards posterior tree space (marked by the red squares) in fewer iterations compared to the STL kernel analyses. Dark blue points show trees sampled during given time window. Light blue dots show trees sampled in previous time windows. Grey background shows all trees visited during the first 50 million iterations for all runs combined. We refer to Supplementary Figure 8 for a similar visualisation after discarding the first 500 million iterations of the analyses.

Despite the fact that classical multidimensional scaling has been the standard for topological convergence assessment (Smith, 2022; Gao et al., 2026), there are possibly better ways of projecting tree space into lower-dimensional visualisations, even though the additional visualisations in Supplementary Figures S4 and S5 provide confidence that our approach is sufficiently representative for our purposes. Dimensionality reduction techniques like MDS rely on a set of pairwise phylogenetic distances to quantify (dis)similarities between trees. There are however many metrics for doing so (Brusselmans et al., 2024), and there is currently no method that is proven to be consistently superior for phylodynamic purposes. Smith (2022) makes strong theoretical arguments for why RF distances are suboptimal as they are easily saturated by small changes in topology and therefore don't necessarily reflect phylogenetically meaningful differences. The author instead argues for the use of quartet distances (Estabrook et al., 1985), which they show can better identify trends in tree space and map more faithfully to low-dimensionality projections. However, the use of quartet distances scales poorly with data set size. The largest tree analysed by Smith (2022) contained 88 taxa, which is substantially smaller than what can be expected in modern genomic epidemiology and the data sets we analysed in this study. Trees larger than 477 taxa contain more quartets than can be represented by CRAN R's signed 32-bit integers (R Core Team, 2025), and we observed computational difficulties well below that point. Other software, like the Python package tqDist (Sand et al., 2014), might be able to remedy this, as their experiments show functionality for trees up to 10 000 taxa. Solving how tree space should be visualised is beyond the scope of this study, but the consistency in our results when comparing the distance-based tree space visualisations on the one hand, and the statistic-based convergence and mixing diagnostic on the other, provides confidence that the chosen methods are appropriate.

In this study, we quantified kernel mix performance in terms of numbers of iterations needed to reach certain diagnostic thresholds, but so far omitted the question of whether the classic kernel mix and the STL kernel mix perform the same number of MCMC iterations over a given period of real-world ("wall-clock") time. Supplementary Figure S15 shows the computational speeds of the MCMC iterations for all replicate analyses for the three data sets studied here. The STL kernel mix is on average 8.1% slower than the classic kernel mix for the HIV data, has practically indistinguishable performance for the ZIKV data, and is 3.1% faster than the classic kernel mix for the BEAR data. This may suggest a positive association between data set size and added benefit in terms of speed from using STL, but additional work is needed to confirm. Importantly, this shows that the improved posterior tree space exploration aspects by the STL kernel mix do not come with a computational penalty that would undo the convergence and split-frequency ESS accumulation benefits.

An important caveat regarding our analyses is that we fixed the relative weighting schemes of all transition kernels to the current default settings implemented in BEAUti X (v1.10.5) (Baele et al., 2025) for both the classic and STL kernel mixes, while fixing all substitution model parameters to their posterior mean values from an initial data-exploration analysis. Our aim was thus to compare the relative performances of the two kernel mixes for exploring tree space by minimising the influence of outside factors. Finding an optimal weighting scheme for the individual transition kernels that make up the kernel mixes falls beyond the scope of the present study. The current STL weighting scheme assigns more relative weight to its transition kernels proportionally to the size of the data set being studied (see Table 1), while the classic transition kernels are assigned a fixed weight. Finding an optimal relative kernel weighting scheme will involve a substantial amount of experiments and may turn out to be model and data set dependent, and is the subject of ongoing work.

Finally, the current STL implementation uses a simple truncated half-normal distribution for the distance kernel $\delta$, but any strictly positive distribution would work, and there are possibly better options. Bactrian distributions (Yang and Rodríguez, 2013) could be explored in order to simultaneously accommodate conservative and bold (potentially mode-jumping) proposals. Another strategy is to run multiple instances of STL in parallel with different distance kernels, such that each could “specialise” in a specific type of movement through tree space. We also believe the added benefit of STL lies not just in its demonstrated performance increases, but also in its modular nature, allowing for further improvements down the line. Currently, STL picks its node to prune completely at random. A targeted approach favoring nodes whose parent branch contributes highly to the trees’ total parsimony score could increase efficiency by decreasing correlation between subsequent trees and thus improving effective sample size per unit time (Bouckaert et al., 2025). A final point of attention is the optimal acceptance probability of adaptive tree kernels. We have here employed the commonly-used value of 0.234, but formal experiments might reveal different targets may offer superior performance depending on the model and data set (Bédard, 2008; Li et al., 2025).

# Data availability

All *BEAST X* XML files used in the analyses can be found freely on:
https://github.com/Brusselmans-Marius/SubTreeLeap-Data.git

For reviewers:
http://datadryad.org/share/LINK_NOT_FOR_PUBLICATION/QGdf8a60laBjrSX0pgTLAJevx9i0Dif8tuSqND6NhR0

# Acknowledgements

We would like to thank Dootika Vats for clarifications on multivariate ESS. MAS is partly supported by NIH grants U19 AI135995, R01 AI153044 and AI162611. JG also acknowledges support from NIH grant R01 AI162611. GB acknowledges support from the Research Foundation - Flanders ("Fonds voor Wetenschappelijk Onderzoek - Vlaanderen," G098321N) and from the European Union Horizon 2023 RIA project LEAPS (grant agreement no. 101094685). GB and MB acknowledge support from the DURABLE EU4Health project 02/2023-01/2027, which is co-funded by the European Union (call EU4H-2021-PJ4) under Grant Agreement No. 101102733. Views and opinions expressed are however those of the author(s) only and do not necessarily reflect those of the European Union or the European Health and Digital Executive Agency. Neither the European Union nor the granting authority can be held responsible for them. LMC was partly funded by FAPERJ (Grant No. 260003/013252/2024 and 260003/005679/2023) and CNPq (Grant No. 303625/2026-0).

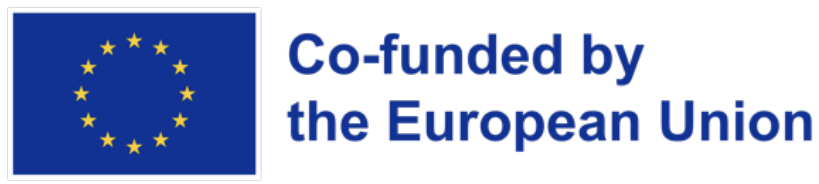

# Supplementary Materials

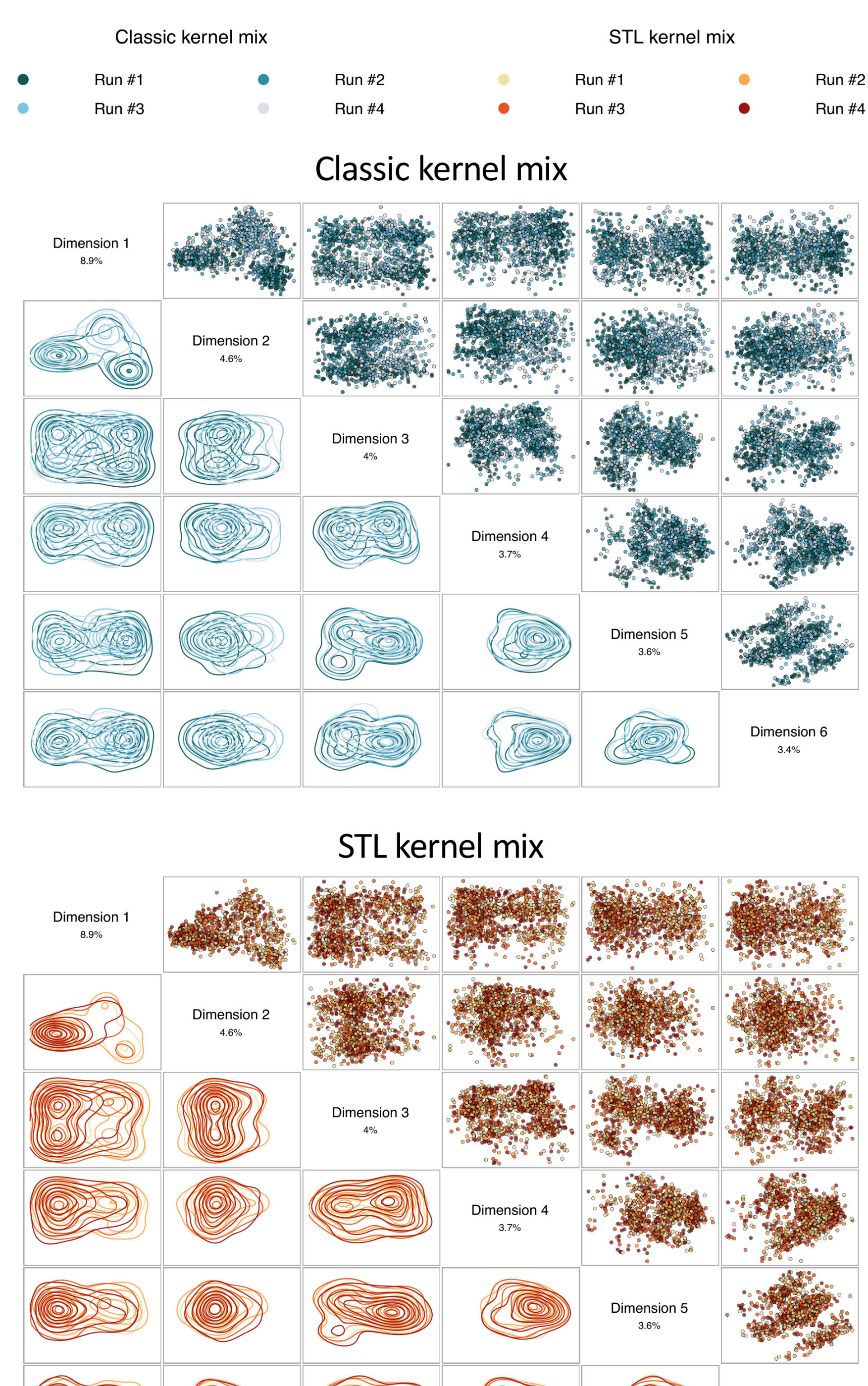

Supplementary Figure S1: **Topological mixing analysis for the HIV data set.** A 6-dimensional MDS performed on RF-distances between 300 equally spaced posterior samples of each run (after burn-in) shows a largely homogenous explored tree space by both kernel types (top: classic kernel mix; bottom: STL kernel mix). Above the diagonal are trees MDS coordinates for each tree, below the diagonal are 2D kernel density contour lines of said coordinates. Percentages shown are the proportion of explained variability by each dimension.

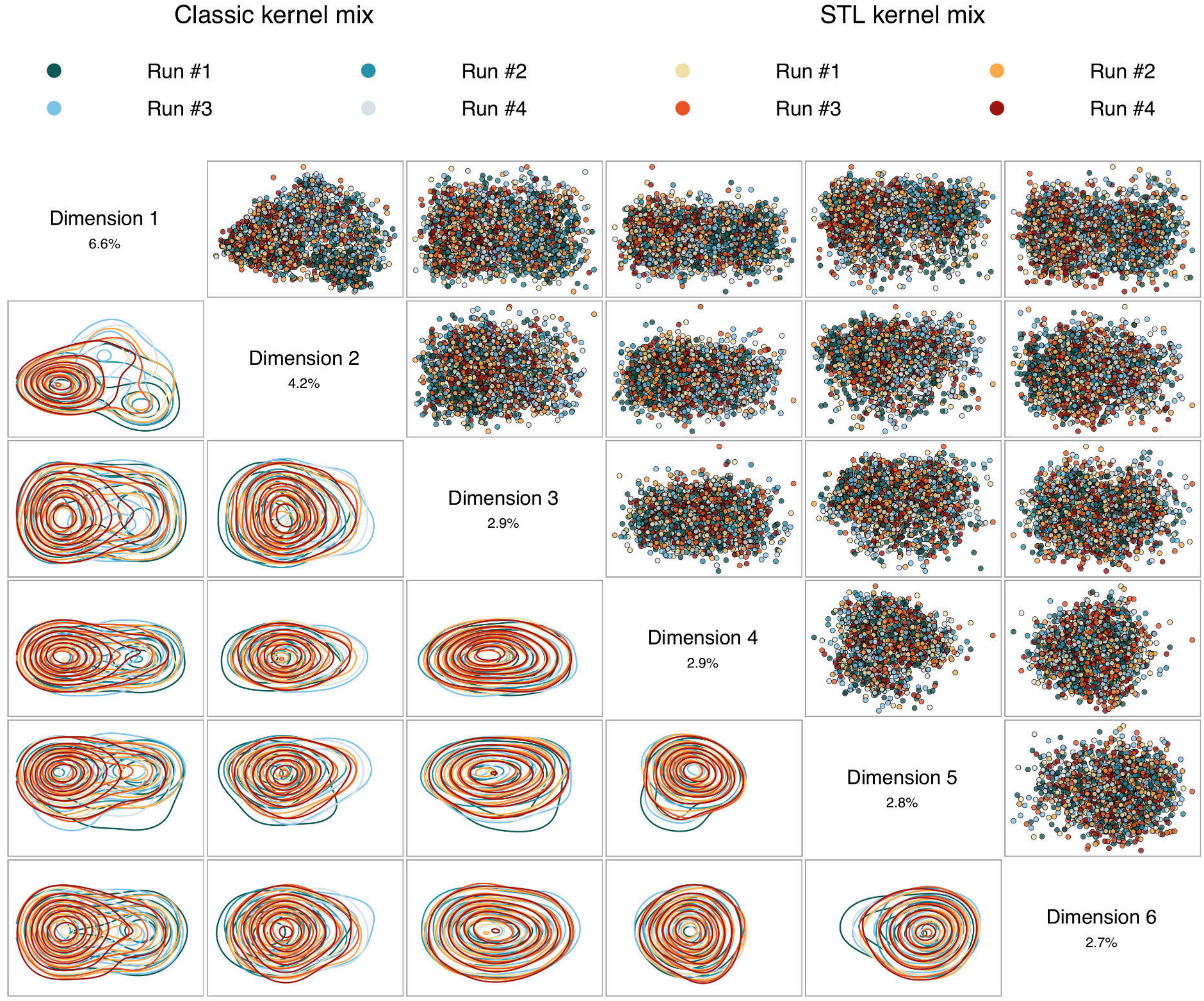

Supplementary Figure S2: **Topological mixing analysis for the HIV data set. (SPR)** A 6-dimensional MDS performed on SPR distances between 300 equally spaced posterior samples of each run (after burn-in) shows a largely homogenous explored tree space by both kernel types. Minor differences between kernel types are due to the classic kernel mix tending to remain in the same posterior tree region, as shown in Figure 4. Above the diagonal are trees MDS coordinates for each tree, below the diagonal are 2D kernel density contour lines of said coordinates. Percentages shown are the proportion of explained variability by each dimension.

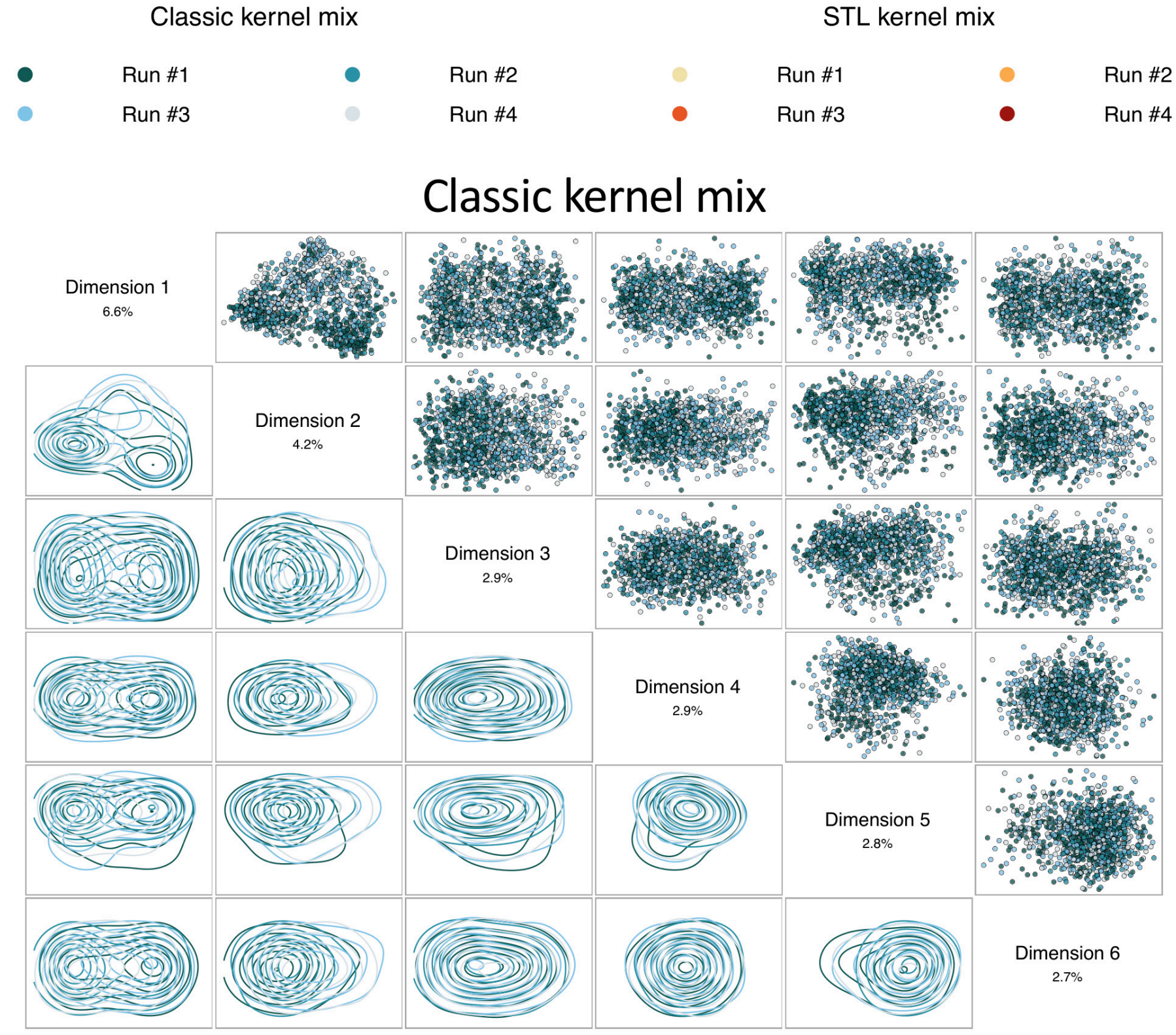

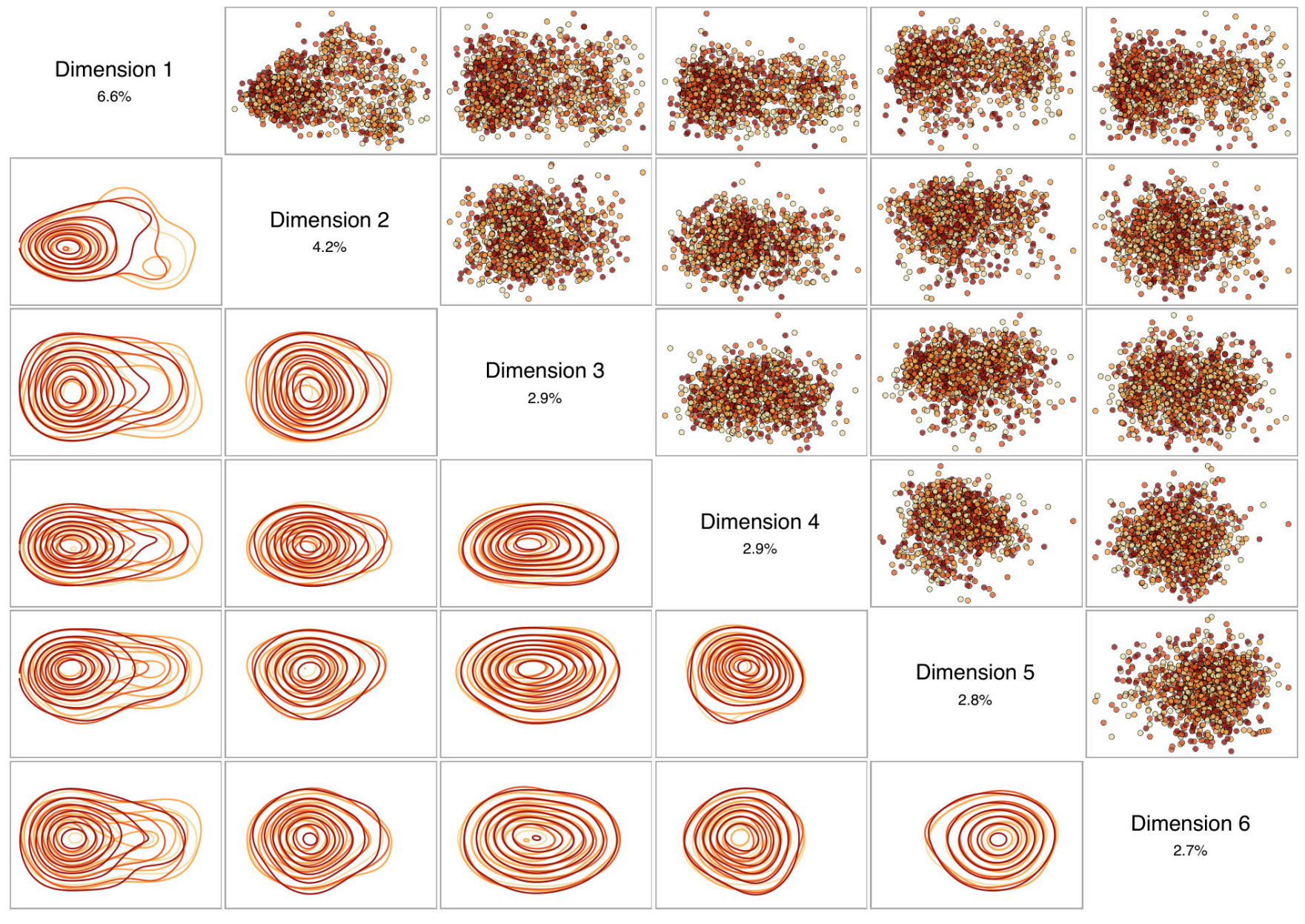

Supplementary Figure S3: **Topological mixing analysis for the HIV data set. (SPR)** A 6-dimensional MDS performed on SPR distances between 300 equally spaced posterior samples of each run (after burn-in) shows a largely homogenous explored tree space by both kernel types (top: classic kernel mix; bottom: STL kernel mix). Minor differences between kernel types are due to the classic kernel mix tending to remain in the same posterior tree region, as shown in Figure 4. Above the diagonal are trees MDS coordinates for each tree, below the diagonal are 2D kernel density contour lines of said coordinates. Percentages shown are the proportion of explained variability by each dimension.

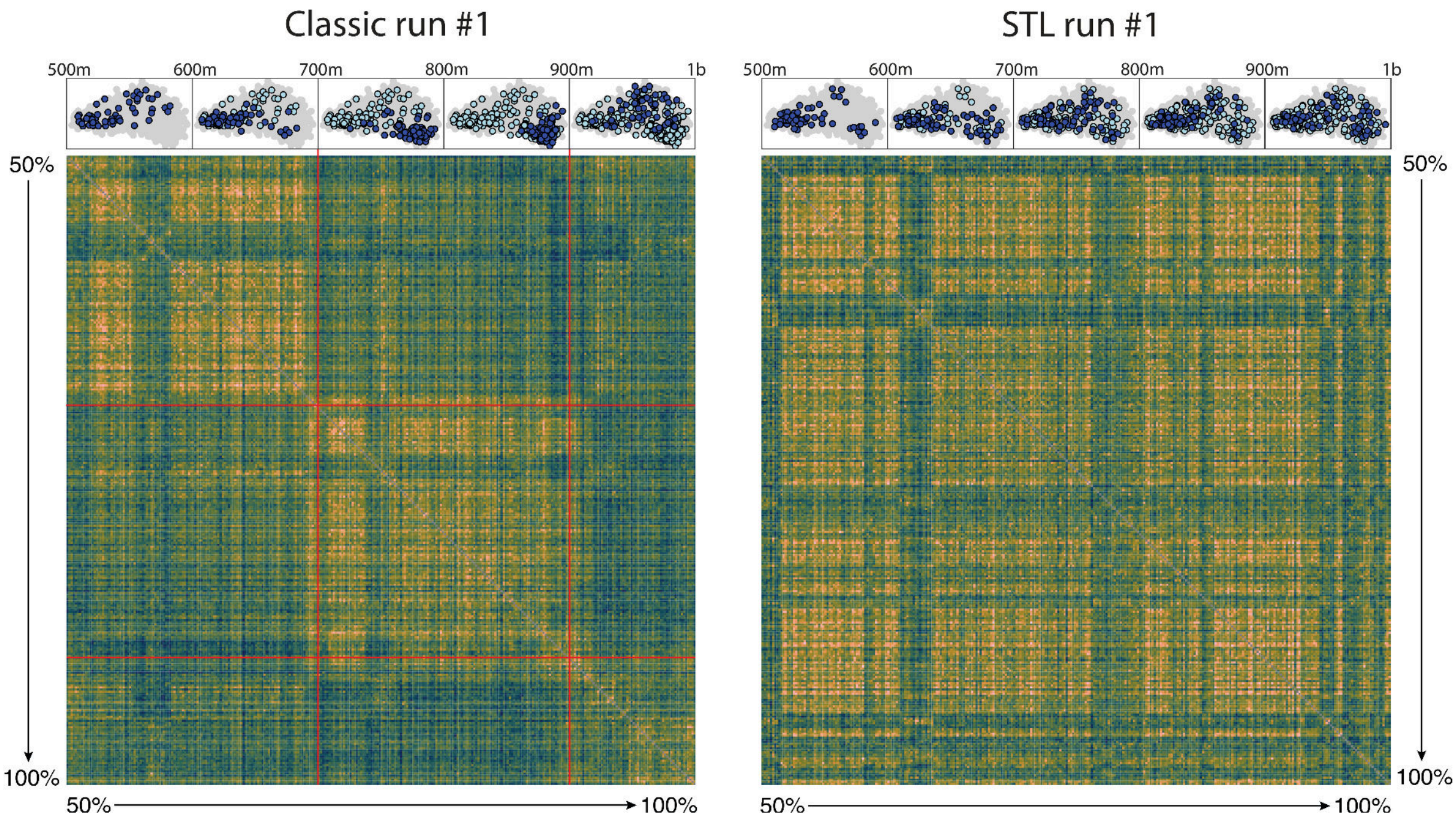

Supplementary Figure S4: **Pairwise RF distances between two post-burn-in tree samples using a classic kernel and STL kernel mix for the HIV data set.** Each heat map shows pairwise RF distances between 300 equally spaced trees after burn-in, corresponding to the posterior trees seen in the tree space exploration progression on top (see Figure 4 - Classic run #1 & STL run #1). The slower mixing of the classic kernel mix is apparent in the larger “clumps” of more similar trees as seen along the diagonal (delineated by red lines) of the heat map, corresponding to the Markov chain exploring the left and right sides respectively of the MDS projected tree space. The STL kernel mix shows a much more noisy / heterogeneous heat map, as expected from better mixing where the Markov chain does not remain stuck in a specific part of posterior tree space for (as) long.

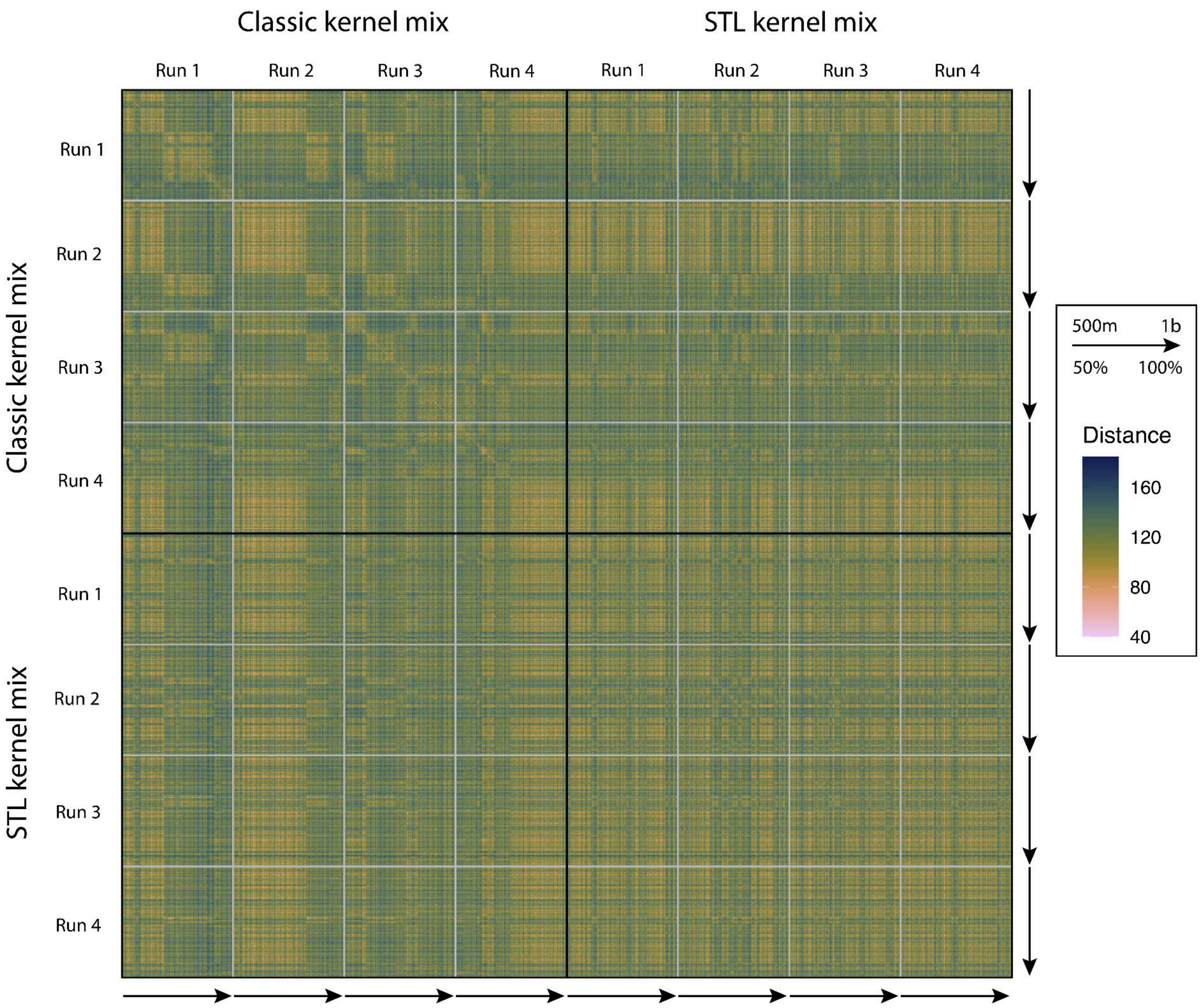

Supplementary Figure S5: **Pairwise RF distances between all post-burn-in tree samples using a classic kernel and STL kernel mix for the HIV data set.** Black lines delineate tree transition kernels, grey lines delineate individual MCMC analyses. Arrows show start and end of each observation period (50% to 100% of each MCMC analysis, corresponding to 500 million iterations post burn-in). The top left/bottom right square do not show lower average distances than the top right/bottom left square, which would be expected if there was a systematic difference between tree spaces explored by the classic and STL kernel mixes (leading to larger between-kernel-mix distances than within-kernel-mix distances). This supports our finding that the classic kernel mix and the STL kernel mix explore the same posterior tree space.

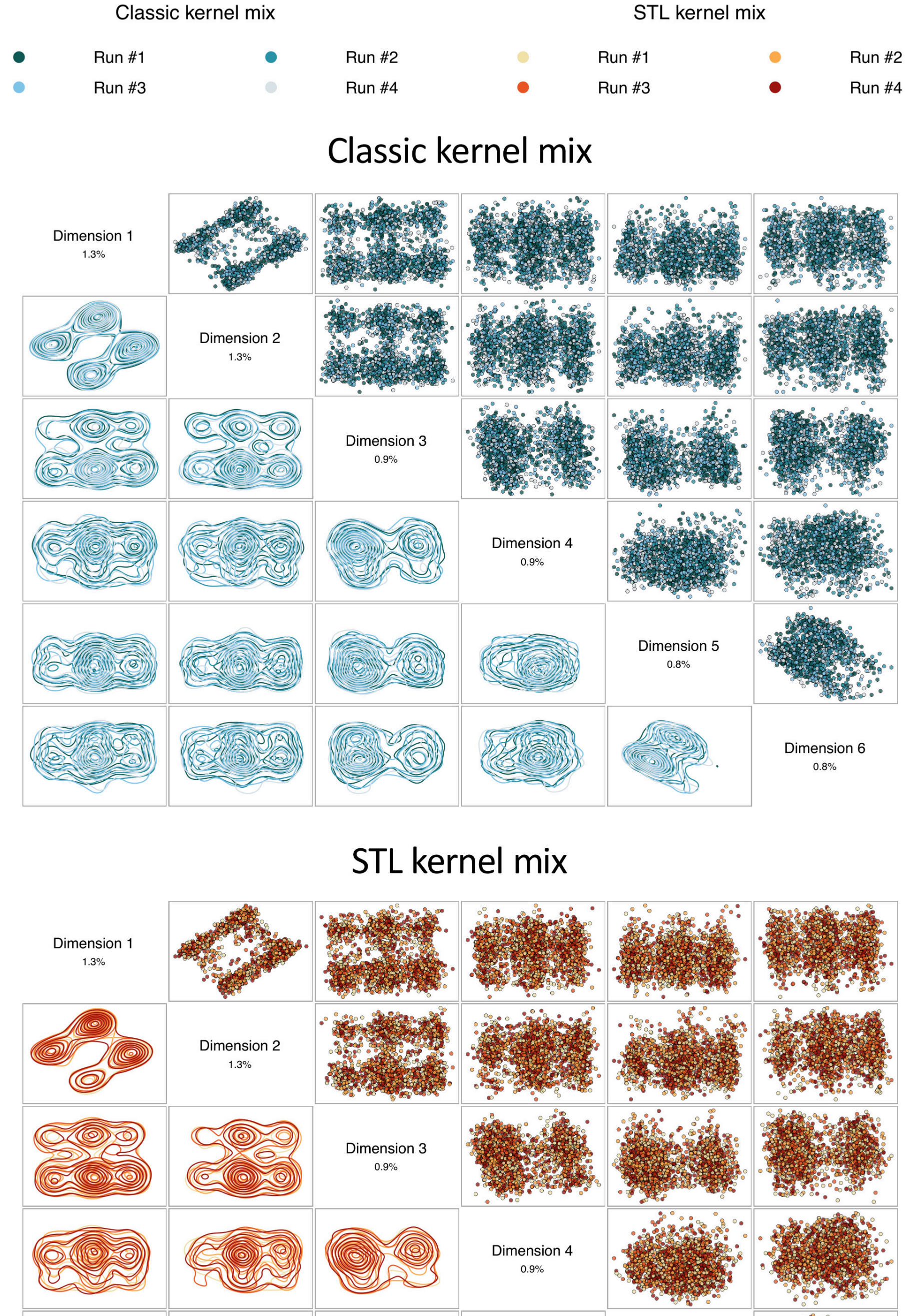

Supplementary Figure S6: **Topological mixing analysis for the ZIKV data set.** A 6-dimensional MDS performed on RF-distances between 300 equally spaced posterior samples of each run (after burn-in) shows a largely homogenous explored tree space by both kernel types (top: classic kernel mix; bottom: STL kernel mix). Above the diagonal are trees MDS coordinates for each tree, below the diagonal are 2D kernel density contour lines of said coordinates. Percentages shown are the proportion of explained variability by each dimension.

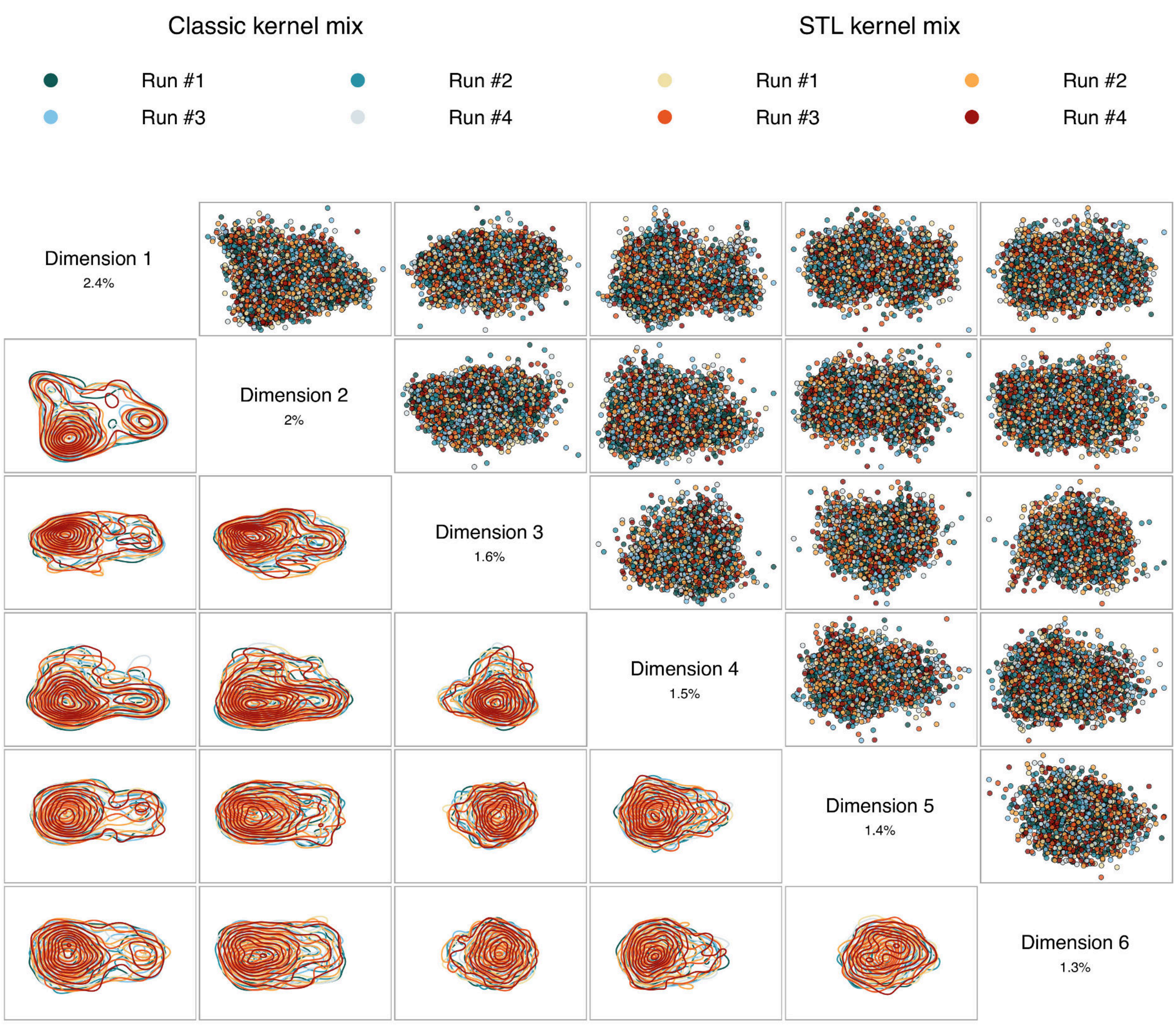

Supplementary Figure S7: **Topological mixing analysis for the ZIKV data set. (SPR)** A 6-dimensional MDS performed on SPR distances between 300 equally spaced posterior samples of each run (after burn-in) shows a largely homogenous explored tree space by both kernel types. Above the diagonal are trees MDS coordinates for each tree, below the diagonal are 2D kernel density contour lines of said coordinates. Percentages shown are the proportion of explained variability by each dimension.

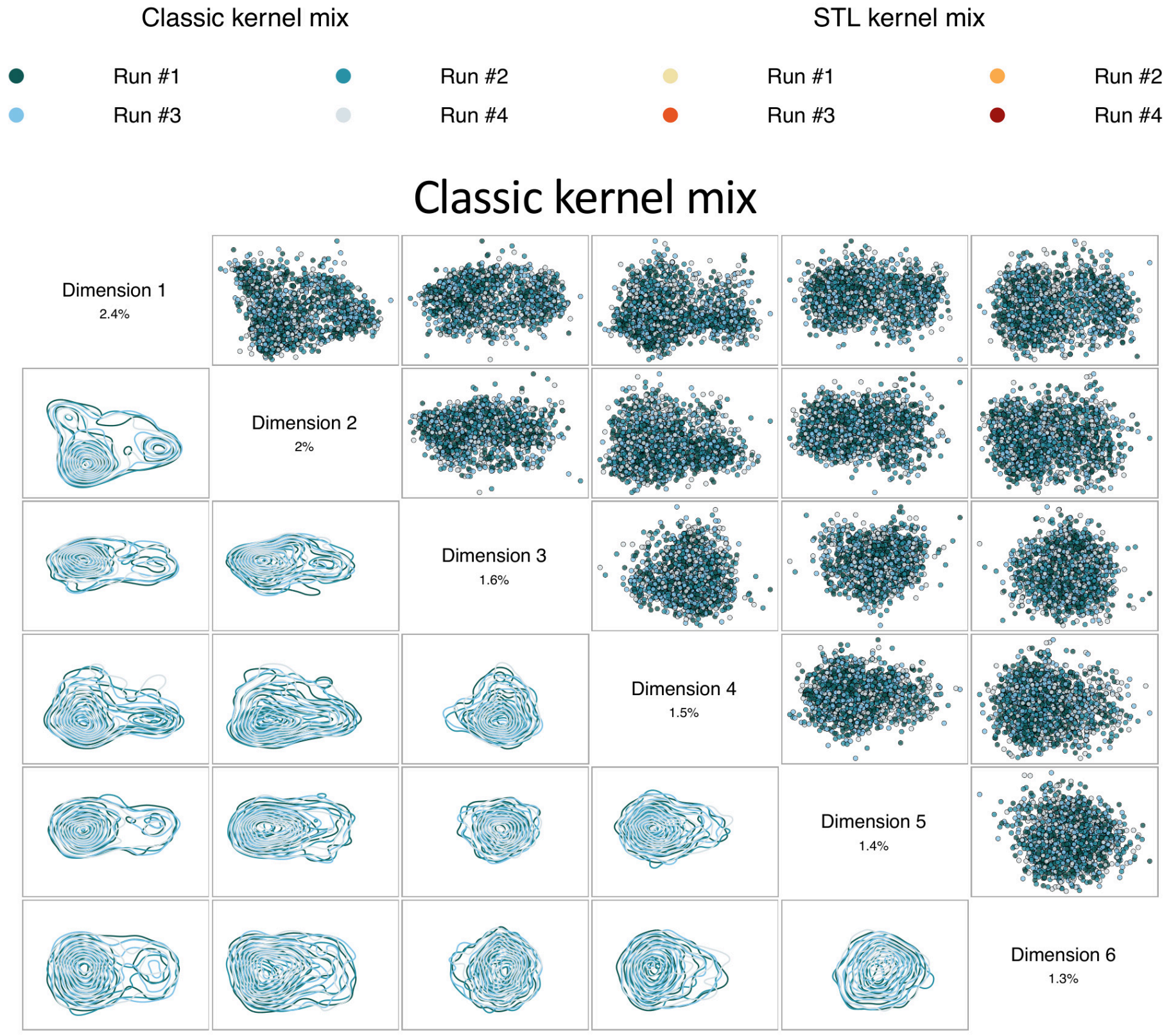

STL kernel mix

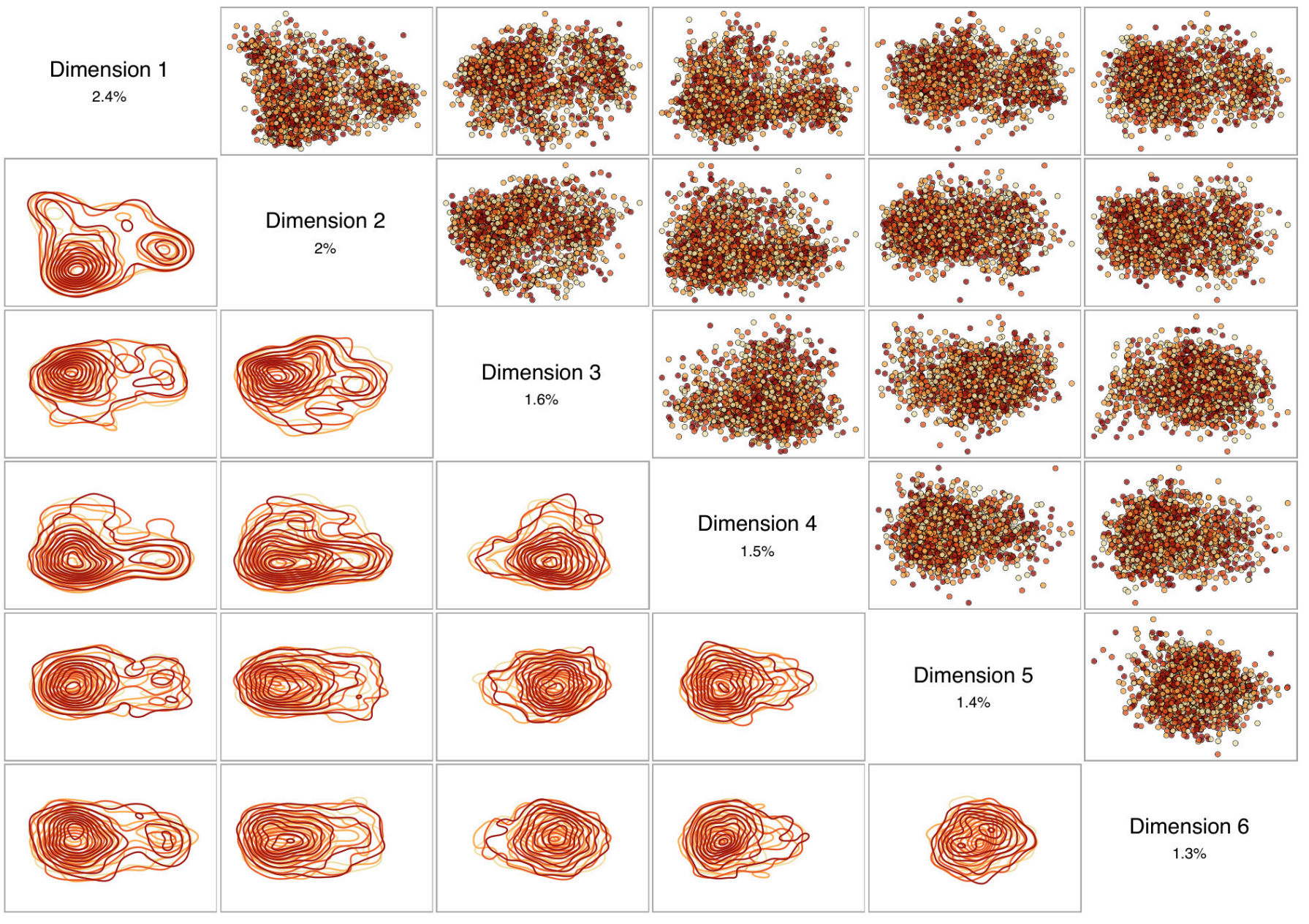

Supplementary Figure S8: **Topological mixing analysis for the ZIKV data set. (SPR)** A 6-dimensional MDS performed on SPR distances between 300 equally spaced posterior samples of each run (after burn-in) shows a largely homogenous explored tree space by both kernel types (top: classic kernel mix; bottom: STL kernel mix). Above the diagonal are trees MDS coordinates for each tree, below the diagonal are 2D kernel density contour lines of said coordinates. Percentages shown are the proportion of explained variability by each dimension.

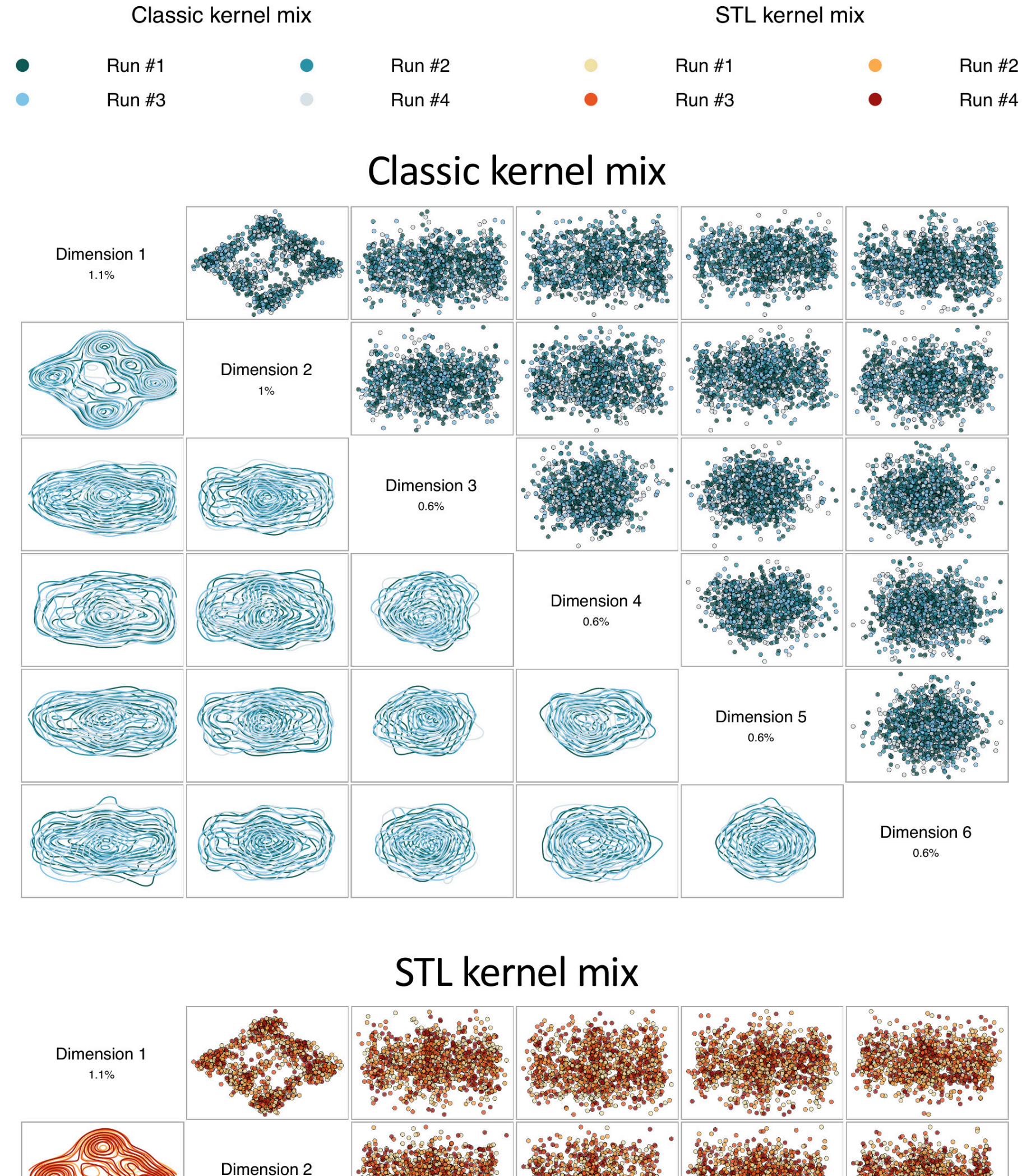

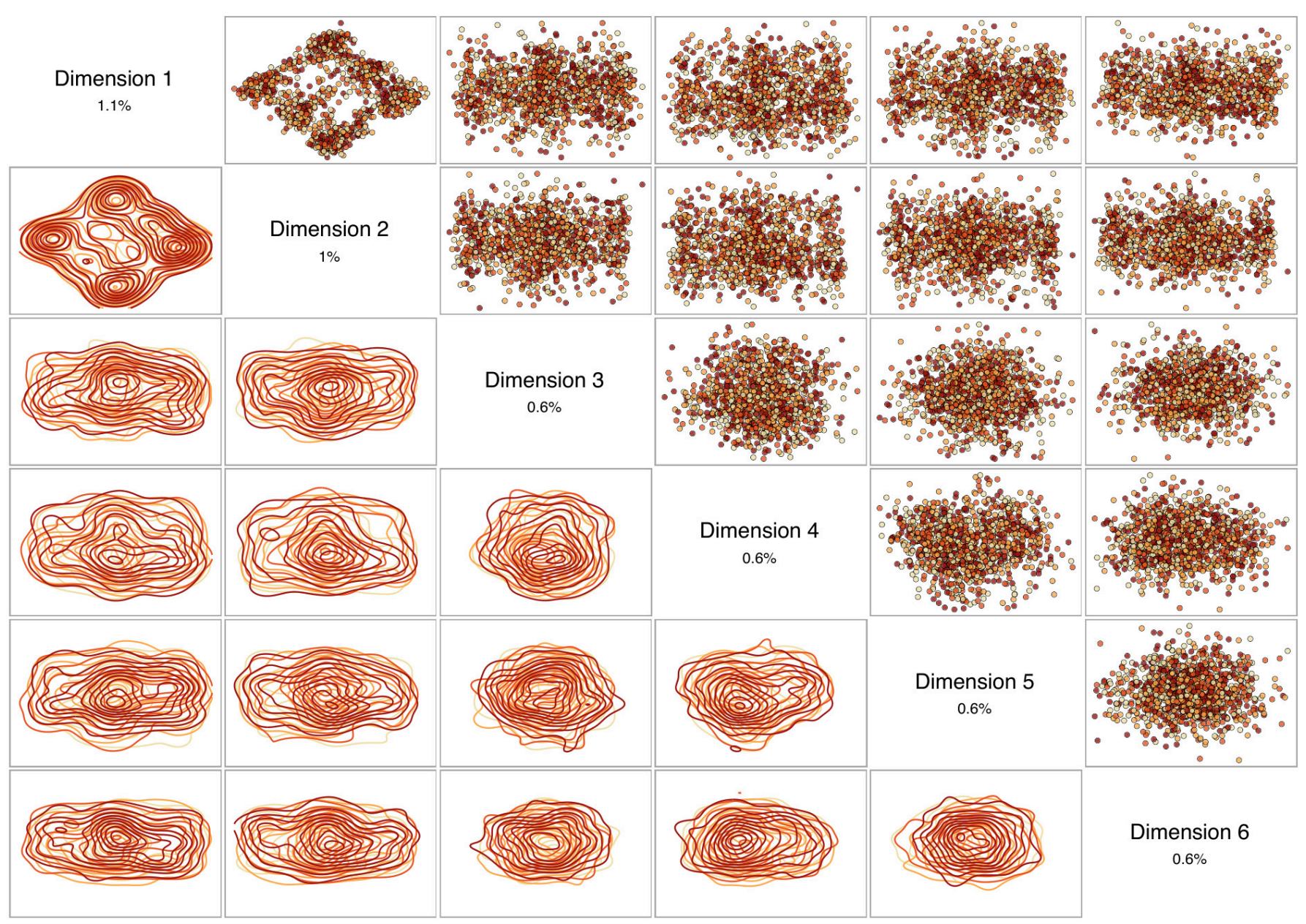

Supplementary Figure S9: **Topological mixing analysis for the BEAR data set.** A 6-dimensional MDS performed on RF-distances between 300 equally spaced posterior samples of each run (after burn-in) shows a largely homogenous explored tree space by both kernel types (top: classic kernel mix; bottom: STL kernel mix). Above the diagonal are trees MDS coordinates for each tree, below the diagonal are 2D kernel density contour lines of said coordinates. Percentages shown are the proportion of explained variability by each dimension.

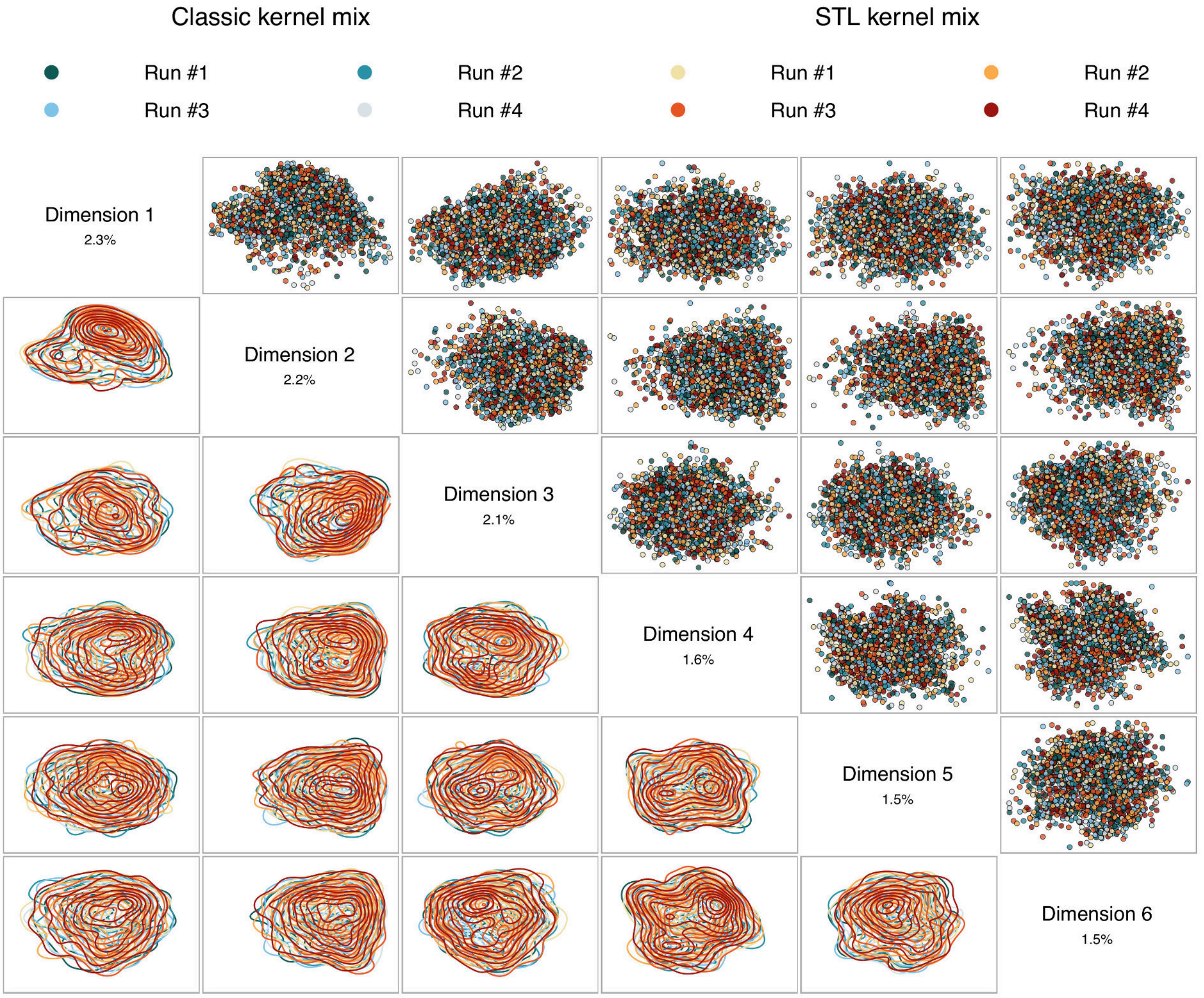

Supplementary Figure S10: **Topological mixing analysis for the BEAR data set. (SPR)** A 6-dimensional MDS performed on SPR distances between 300 equally spaced posterior samples of each run (after burn-in) shows a largely homogenous explored tree space by both kernel types. Above the diagonal are trees MDS coordinates for each tree, below the diagonal are 2D kernel density contour lines of said coordinates. Percentages shown are the proportion of explained variability by each dimension.

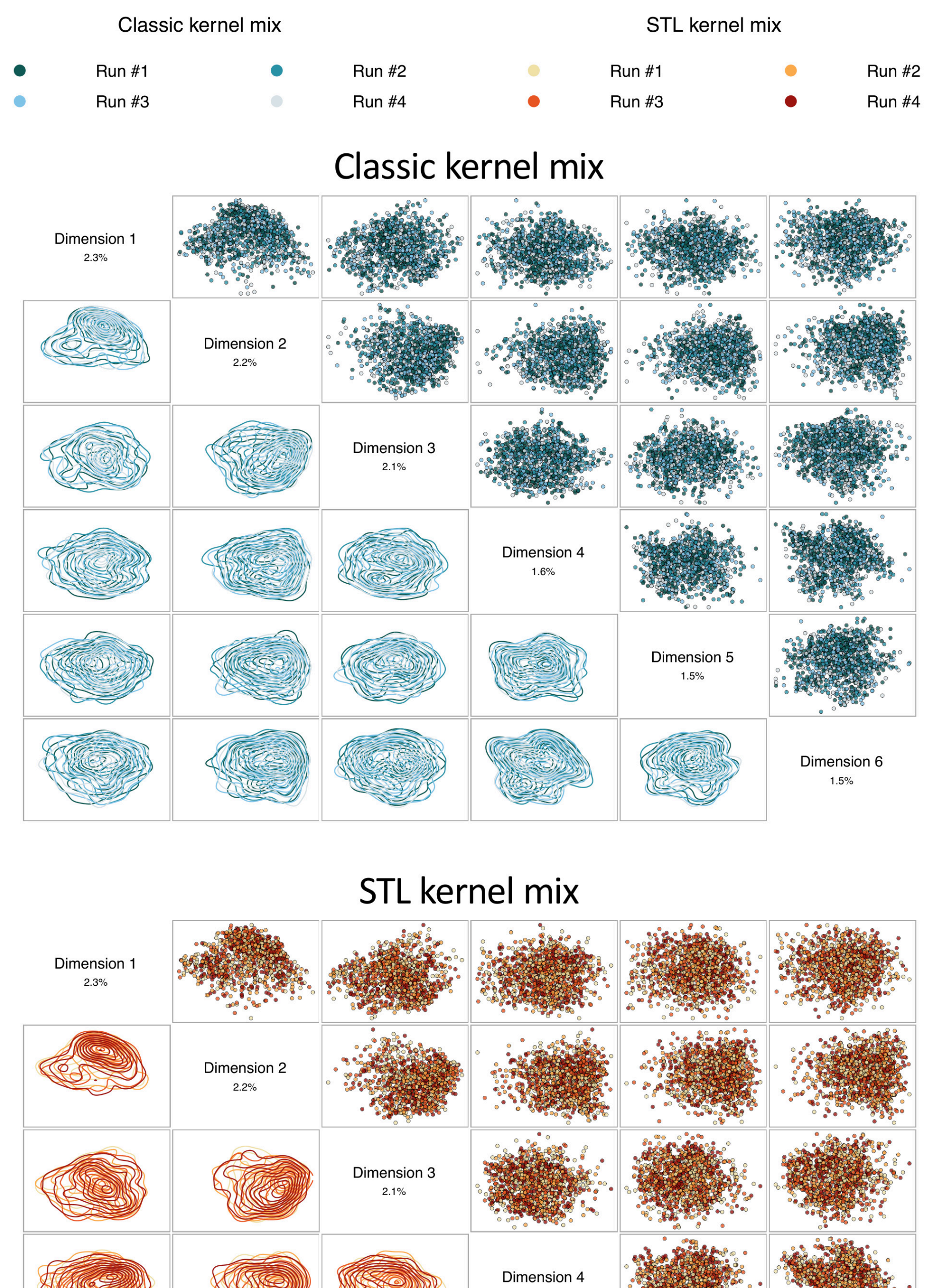

Supplementary Figure S11: **Topological mixing analysis for the BEAR data set. (SPR)** A 6-dimensional MDS performed on SPR distances between 300 equally spaced posterior samples of each run (after burn-in) shows a largely homogenous explored tree space by both kernel types (top: classic kernel mix; bottom: STL kernel mix). Above the diagonal are trees MDS coordinates for each tree, below the diagonal are 2D kernel density contour lines of said coordinates. Percentages shown are the proportion of explained variability by each dimension.

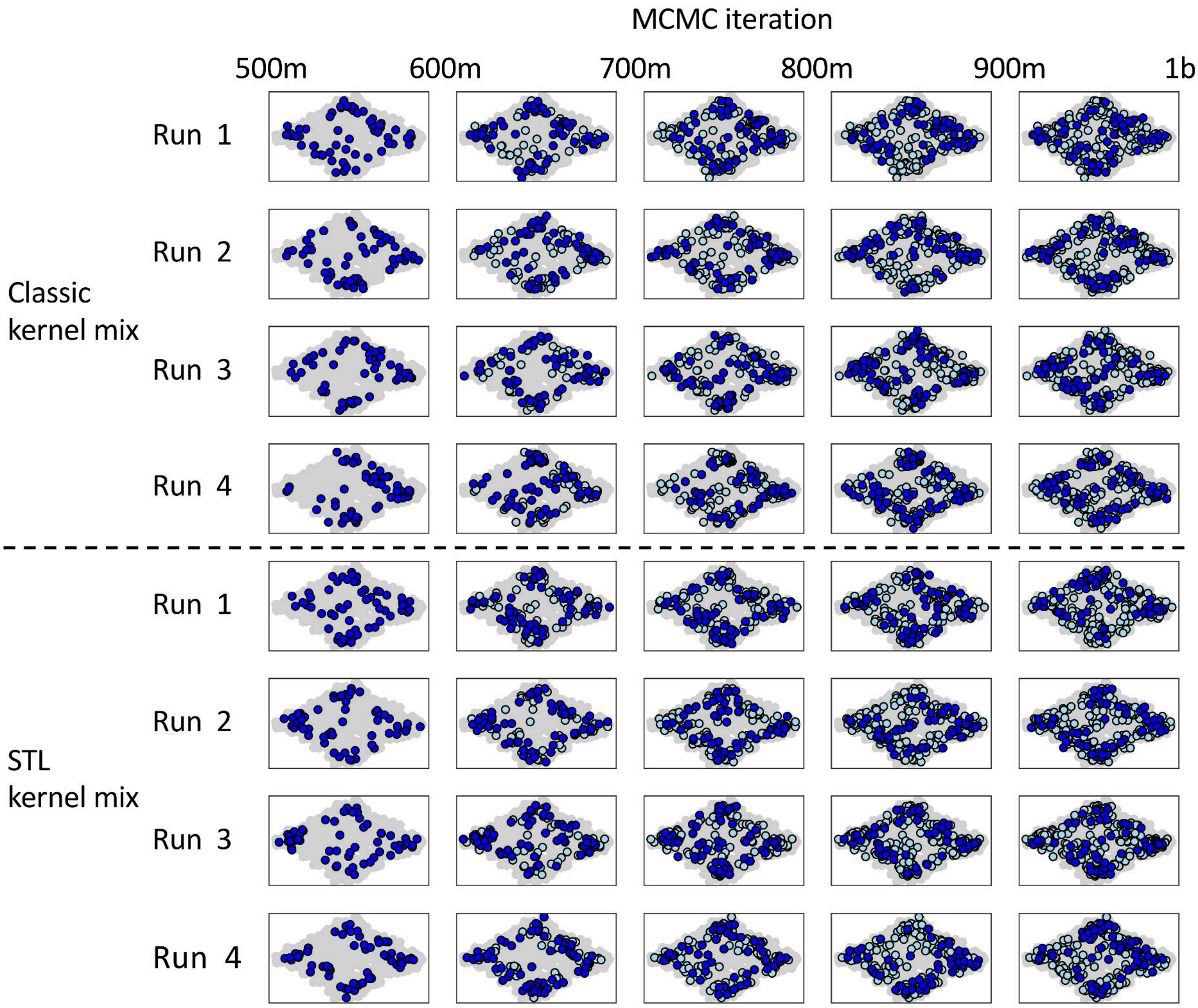

Supplementary Figure S12: **Dimensions 1 and 2 of RF distance-based PCoA posterior tree space for the BEAR data set.** After discarding the first 500 million iterations as burn-in, this figure shows how both classic and STL kernel mixes explore posterior tree space throughout the following 500 million iterations. Dark blue points show posterior trees sampled during a given window of 100 million iterations. Light blue dots show all trees sampled up to that point. Grey background dots show the entire posterior tree space for the post-burn-in 500 million iterations for all runs combined. Both kernel mixes show temporally consistent homogeneous explorations of posterior tree space and equally frequent transitions between the modes in the RF distance-based MDS projection.

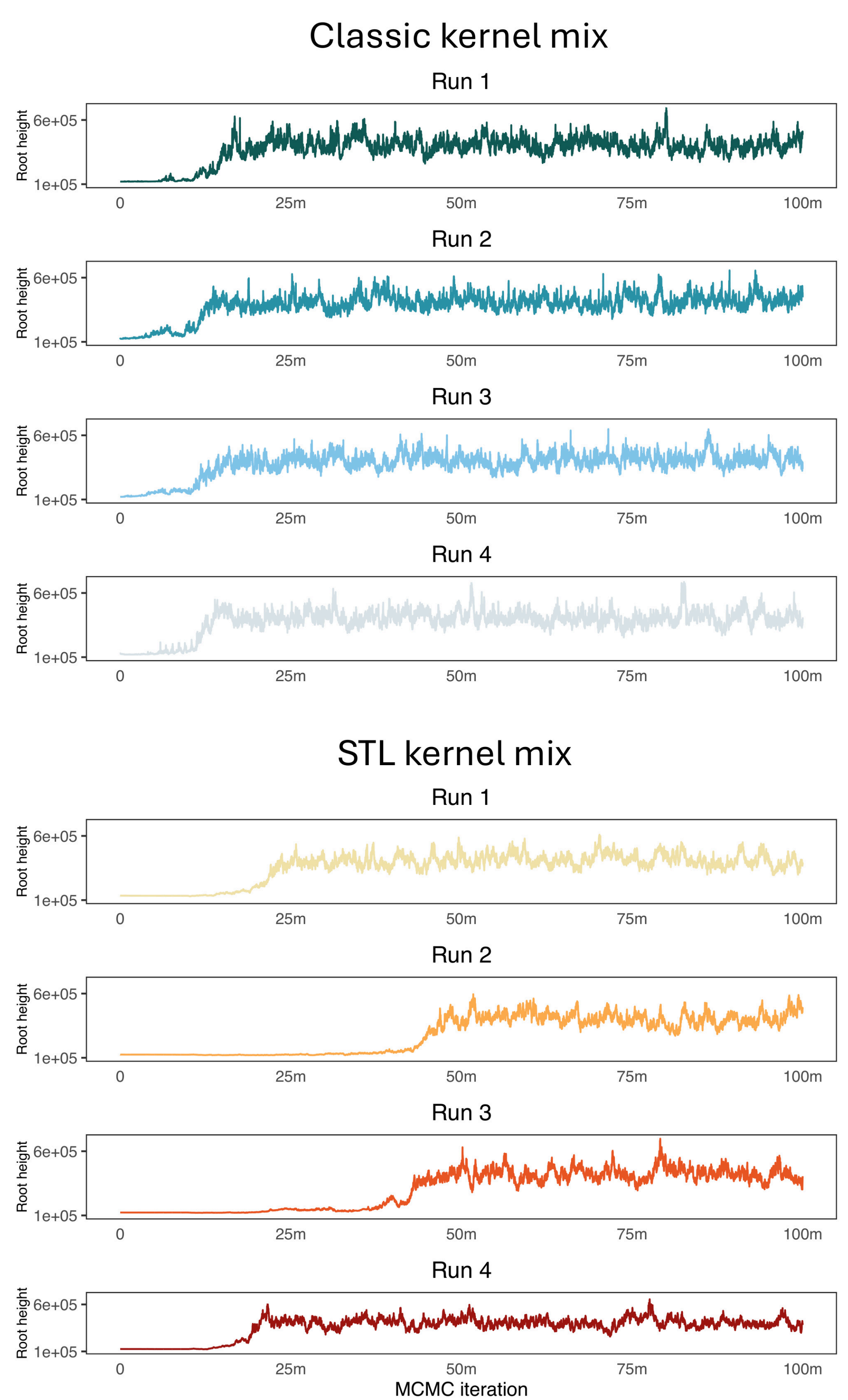

Supplementary Figure S13: **Convergence of the root height for the BEAR data set.** Traces for the root height parameter for the first 100 million iterations of each replicate analysis show that the root height consistently converges faster to its posterior distribution using the classic kernel mix compared to the STL kernel mix. This is consistent with the ASDSF curves for the BEAR data set as shown in Figure 7A.

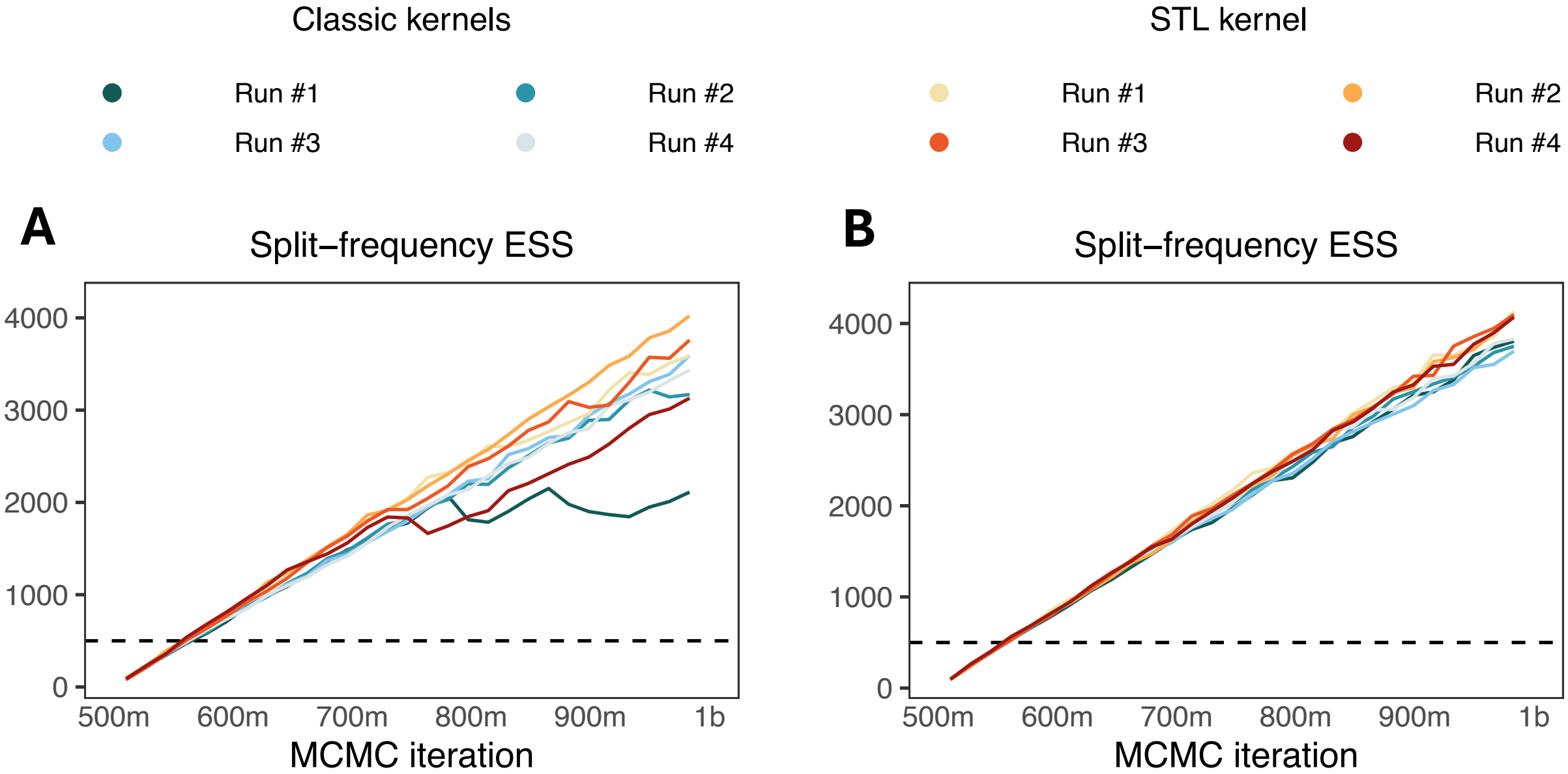

Supplementary Figure S14: **Extended split-frequency ESS for ZIKA en BEAR data for latter 500 million MCMC iterations**. Both data sets are further downsampled tenfold. ESS curves are consistent with those throughout the rest of the study as they show a slightly faster growth for analyses using the STL kernel mix as opposed to the classic kernel mix. In the ZIKA analysis, STL run 4 shows a slump the curve, coinciding with a moment of being "stuck" in tree-space, before rectifying and regaining its initial slope. Classic run 1 shows two such slumps.

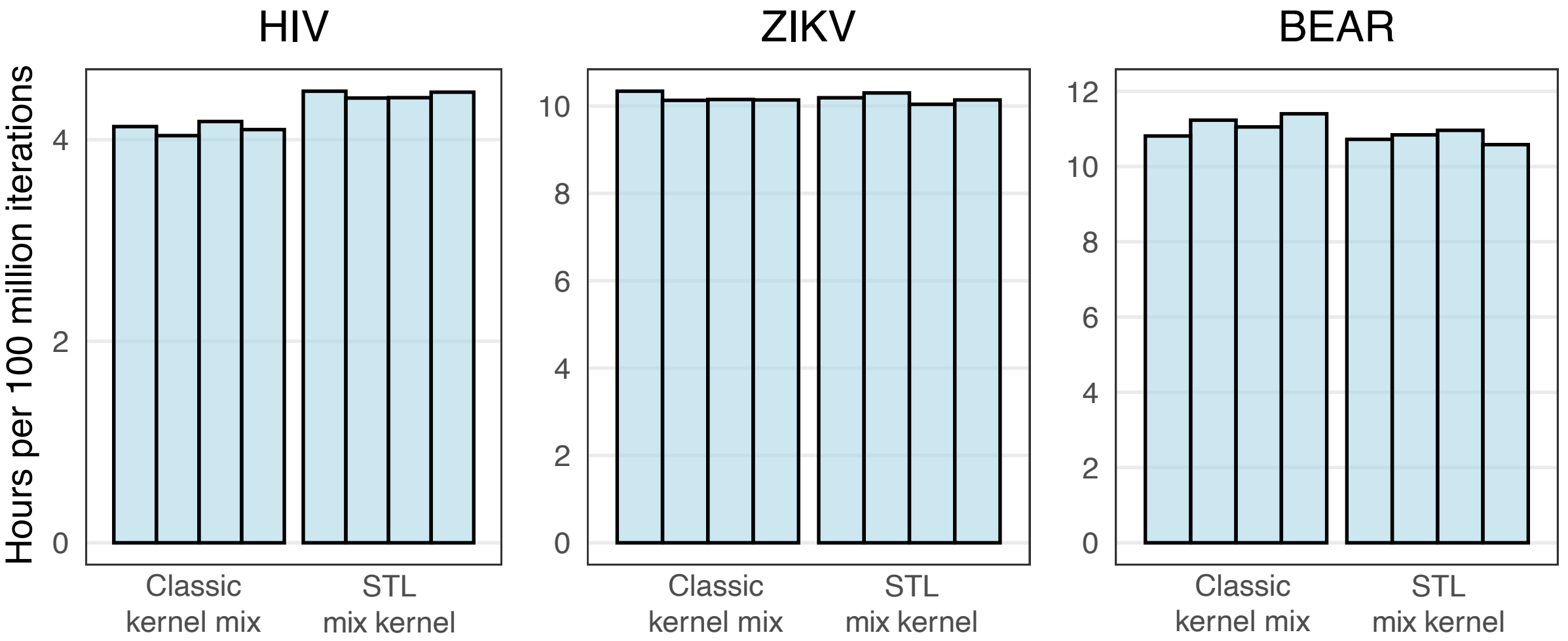

Supplementary Figure S15: **Wall-clock MCMC computing time by kernel type.** For each data set, the time per 100 million iterations for each of the four replicate analysis for each kernel type, as performed in BEAST X (v1.10.5) (Baele et al., 2025) using BEAGLE 4 (Gangavarapu et al., 2026). Differences in computing time are small, with the STL mix on average being 8.1% slower than the classic kernel mix for the HIV data, practically indistinguishable for the ZIKV data, and 3.1% faster for the BEAR data. These computations were run on a system equipped with 2 × Intel Xeon Gold 6140 CPUs and an NVIDIA Tesla P100 SXM2 GPU (16 GiB HBM2).

# Appendix A: Proofs

We begin by setting out some additional mathematical notation that will ease presentation of the more technical results. We go on to provide proofs of Theorem 1 and auxiliary results.

## Additional notation

A rooted binary tree $t \in \mathbb{T}$ on $n$ taxa is a graph $G(\boldsymbol{V}_t, \boldsymbol{E}_t)$ with $2n-2$ edges, $n-1$ internal nodes and $n$ leaf/external nodes, also called taxa – making up a total of $2n-1$ nodes. Each vertex (node) $v \in \boldsymbol{V}_t$ has degree 3, except for a special *root* internal node, denoted $\rho$, which has degree 2. The set $\boldsymbol{V}_t$ has a partial ordering, defined as follows: $u \preceq v$ if there is a unique simple path from the root $\rho$ to $v$ through $u$, in which case we say $u$ is an *ancestor* of $v$. Denote $\boldsymbol{I}_t = \{x : x \preceq y,\ x, y \in \boldsymbol{V}_t\} \subset \boldsymbol{V}_t$ as the set of interior nodes of $t$. The root node $\rho$ is then the ancestor of all nodes and the smallest element of the ordering imposed by $\preceq$. The set $\boldsymbol{C}_t = \boldsymbol{V}_t \setminus \boldsymbol{I}_t$ is the set of exterior nodes (taxa) of $t$.

Let $X$ be a non-empty set of labels, $\phi : X \to \boldsymbol{V}_t$ be a bijective map and $h : \boldsymbol{I}_t \to \{1, 2, \ldots, |\boldsymbol{I}_t|\}$ an (injective) *ranking* function such that $u \preceq v$ implies $h(u) \leq h(v)$ for all $v, u \in \boldsymbol{I}_t$. Notice that $h(u) = h(v)$ implies either $u \equiv v$ or $u, v \in \boldsymbol{C}_t$. A *ranked rooted tree* is an object $t = (\boldsymbol{V}_t, \boldsymbol{E}_t, \rho, \phi, h),\ t \in \mathbb{F}$. We can supplement $t$ with a set of edge (branch) lengths $\boldsymbol{b} = \{b_1, b_2, \ldots, b_{2n-2}\}, \boldsymbol{b} \in \boldsymbol{B} \subseteq \mathbb{R}_+^{2n-2}$, creating a *fully-ranked rooted phylogeny* in the form of the object $(t, \boldsymbol{b}) = \tau \in \boldsymbol{\Psi}$. For convenience, we will henceforth call $t$ a **topology** and $\tau$ a **phylogeny**.

It is well known that the cardinality of the space of (partially ranked) rooted topologies on $n$ taxa is $R_n = |\mathbb{T}| = n!(n-1)!/2^{n-1}$. Here, however, we are concerned with *fully ranked* phylogenies, specifically those for which the mapping $h$ is a height function that measures node ages in calendar units. If we associate an age in calendar time with each of the $2n-1$ nodes in a rooted binary tree, these can then be ranked and then used to form a poset $\boldsymbol{a} = \{a_1, a_2, \ldots, a_{2n-1}\} \in \boldsymbol{A} \subset \mathbb{R}_+^{2n-1}$. A convenient labelling is to make labels increase with age, such that $i > j$ implies $a_i \geq a_j$; thus, the root node will have label $2n-1$. An edge $e_{i,j}$ with $i > j$ represents an ancestral lineage and node $k$ in $\boldsymbol{I}_t$ corresponds to a *coalescence* event of two ancestral lineages at time $a_k$.

It is convenient to define $\boldsymbol{a}_L$ and $\boldsymbol{a}_I$ as the ages of the leaf and internal nodes, respectively. Also denote $\boldsymbol{a} = (\boldsymbol{a}_I, \boldsymbol{a}_L) \in \boldsymbol{A} \subset \mathbb{R}_+^{2n-1}$. There exists a bijective mapping $D : \boldsymbol{B} \to \boldsymbol{A}$ that maps the branch lengths of a TCP to its node ages. In many phylodynamic applications, taxa are sampled through time, leading to *serially-sampled* data sets (Drummond et al., 2002). Hence, here $\boldsymbol{a}_L$ is fixed (for any $\tau \in \boldsymbol{\Psi}$) as it relates to the data collection process. These sampling patterns are important because they alone impose constraints on the space of phylogenies.

We are now in position to define the set of **intercoalescent intervals** (also called divergence times) as $\boldsymbol{s} = \{s_2, s_3, \ldots, s_n\} \in \boldsymbol{S} \subset R_+^{n-1}$, where $s_i = a_{n-i+1} - a_{n-i}$ for $i = 2, \ldots, n-1$ and $s_n = a_1$. Notice that for trees with tips sampled through time, $\boldsymbol{s}$ will have to be slightly adjusted to include subintervals that correspond to the intervals between either a coalescence or sampling event. As explained by Drummond and Bouckaert (2015), the (infinite) space of time-calibrated phylogenies (TCP) can be composed as $\boldsymbol{\Psi} = \mathbb{F} \times \mathrm{S}$, and it is this space I refer to as **phylogenetic space** throughout the thesis.

Finally, let $k_i$ denote the number of existing lineages in the interval $[a_{i-1}, a_i]$, $\boldsymbol{k} = \{k_2, k_3, \ldots, k_n\} \in \boldsymbol{K} \subset \mathbb{N}^{n-1}$. Defining these quantities – intercoalescent intervals and numbers of lineages – is important in that many prior distributions commonly used in Bayesian phylogenetics are based on coalescent processes and the measure of a phylogeny $\tau$ depends on it only through its coalescent intervals and numbers of lineages, $\boldsymbol{s}(\tau)$. For serially-sampled phylogenies, the number of inter-coalescent intervals, $N_z = |\boldsymbol{s}| = |\boldsymbol{k}|$ is at least $n-1$ and at most[1] $2(n-1)$.

[1]This latter case occurs when there are $n$ distinct sampling times for the $n$ tips and these fall exactly in

## Proofs

**Remark 1.** *The mapping $\psi : (\boldsymbol{T}, \boldsymbol{A}) \longrightarrow (\boldsymbol{S}, \boldsymbol{K})$ is non-injective surjective.*

For a given tree $t$ with associated node times $\boldsymbol{a}$, the intercoalescent times $\boldsymbol{s}$ and lineages through time $\boldsymbol{k}$ can be thought of as summary statistics. It is possible, however, to have pairs of distinct points $\{t, \boldsymbol{a}\}$ and $\{t^*, \boldsymbol{a}\}$ such that $\psi(\{t, \boldsymbol{a}\}) = \psi(\{t^*, \boldsymbol{a}\}) = \{\boldsymbol{s}, \boldsymbol{k}\}$.

To see this, consider the diagram in Supplementary Figure S16.

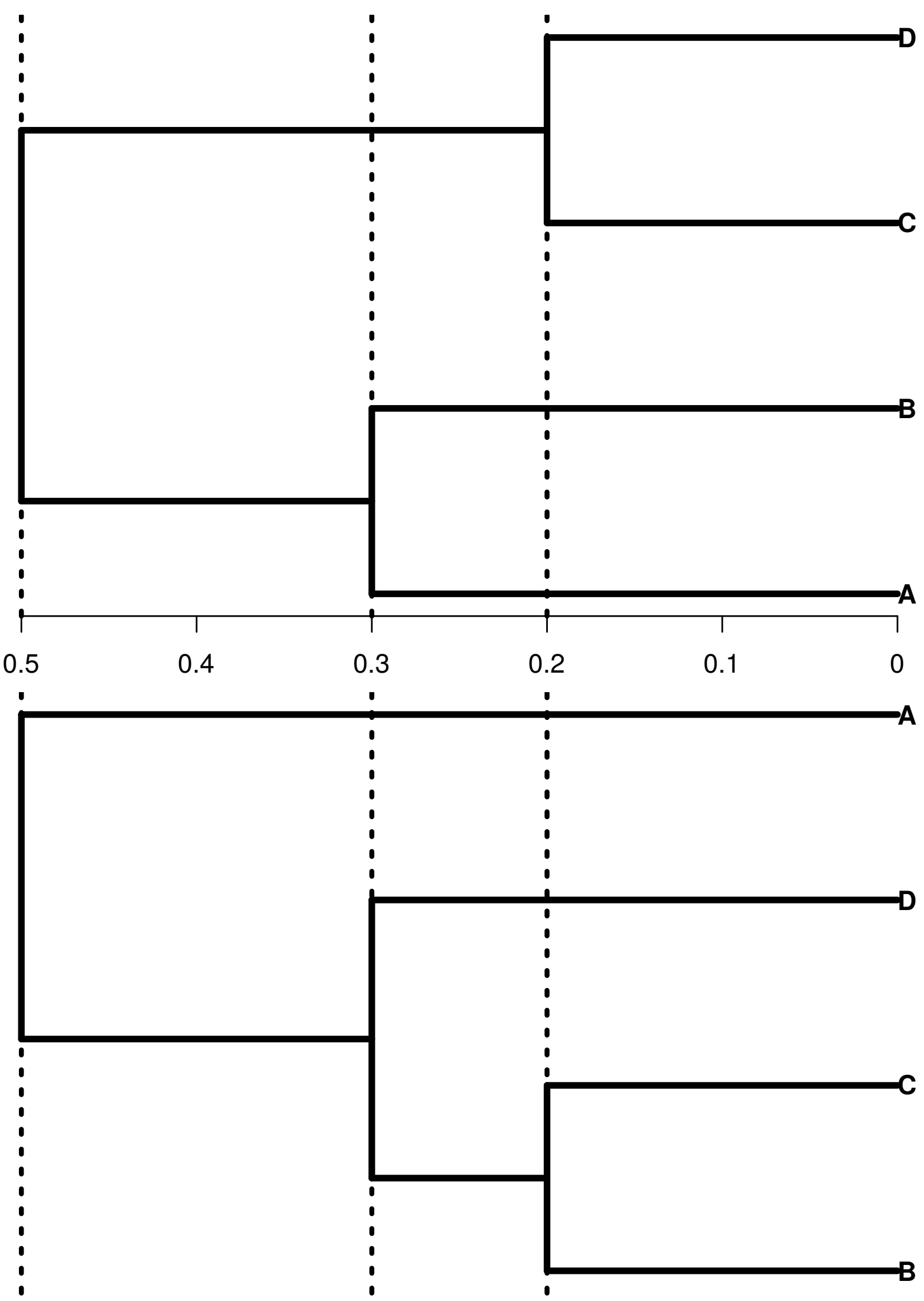

Supplementary Figure S16: **Two distinct trees with the same intercoalescent intervals and numbers of lineages**. In this example, we would have $\boldsymbol{s} = \{0.2, 0.1, 0.2\}$ and $\boldsymbol{k} = \{2, 3, 4\}$.

**Remark 2.** *SubTreeLeap induces an irreducible Markov chain on $\mathbb{F}$.*

*Proof.* First, assume $h(i) > 0 \,\forall i \in V_t$ for any phylogeny $\tau \in \boldsymbol{\Psi}$ with topology $t$. Define $\mathcal{N}(x) = \{u \in \mathbb{F} : d_{\text{SPR}}(u, x) = 1\}$ as the (SPR) *neighbourhood* of $x \in \mathbb{F}$. Irreducibility is equivalent to stating $q_\sigma(y|x) > 0 \,\forall x, y \in \mathbb{F}$.

between sampling dates

Unfortunately, we cannot make this claim directly, because $q_\sigma(y|x) > 0$ only for $y \in \mathcal{N}(x)$ – which is true because $P(x \to y|\delta^\star) > 0$ for $\delta^\star > 0$ and $\kappa(\delta^\star|\sigma) > 0$ for $\sigma > 0$, by construction. However, since the SPR graph is connected (Caceres et al., 2011), it follows that any sequence of topologies $\boldsymbol{X} = \{X^{(0)}, X^{(1)}, \ldots, X^{(N)}\}$ where $X^{(i+1)} \in \mathcal{N}(X^{(i)})$ has positive probability under the transition kernel, establishing irreducibility on the SPR graph and hence on $\mathbb{F}$. □

**Theorem 1.** *SubTreeLeap induces an ergodic Markov chain on* $\boldsymbol{\Psi}$ *with respect to* $\pi$.

*Proof.* We will show that STL is irreducible and aperiodic, which establishes ergodicity (Meyn and Tweedie, 1993; Roberts et al., 2004; Dinh et al., 2017). First, we need to show irreducibility on $\boldsymbol{\Psi}$ by extending the result of Remark 2 to include branch lengths. Suppose there exist $\tau, \tau^\star \in \boldsymbol{\Psi}$ such that $\tau^\star$ cannot be reached from $\tau$ in finitely many STL steps. Following Remark 2, we may assume they have the same topology, *i.e.* $t = t^\star$, and that differences between the two phylogenies lie solely in their branch lengths. This would imply that there exist two phylogenies with the same topology that cannot be transformed into one another in finitely many sliding moves, which is clearly false. Note that STL has a positive probability of producing a sliding move for any node it picks and all nodes (excluding the root) can be picked for any given STL operation. Now, let us show that STL is aperiodic. Let $r_\sigma(\tau)$ be the probability that $t = t'$ and let $A = \{x : r_\sigma(x) > 0\}$. The chain is aperiodic on $\mathbb{F}$ because $p(A) > 0 \; \forall \, A \subset \boldsymbol{\Psi}$ and $r_\sigma(\tau) > 0 \; \forall \, \tau$ – according to Tierney (1994) (pg. 1705) this result can be found in Section 2.4 of Nummelin (1984).

Aperiodicity with respect to branch lengths can be shown using a similar argument to the one used above for irreducibility. We will use the concept of Harris recurrence (Harris, 1956; Chan and Geyer, 1994; Tierney, 1994). First, denote the $n$-th state of the chain by $X^{(n)}$ and define $\kappa_A = \sup\{n \geq 1 : X^{(n)} \in A\}$ for $A \subset \boldsymbol{\Psi}$. Following the definitions and results in Roberts and Rosenthal (2006), for our purposes it suffices to show that $\Pr(\kappa_A < \infty | X^{(0)} = x) = 1$ for all $x \in \boldsymbol{\Phi}$ – since we know that $p(A) > 0$ for all $A$. Again making use of Remark 2, we may restrict attention to a family of (sub)sets $A_t = \{x : d_{\text{SPR}}(x, t) = 0\}$ for some $t \in \mathbb{F}$. Now one can use reasoning by contradiction similarly to what was done above to deduce that $\Pr(\kappa_{A_t} = \infty | X^{(0)} = x) = 0$ for $x \in \boldsymbol{\Phi}$ and for all $t$. If $d_{\text{SPR}}(x, t) = 0$ we use the sliding move argument above, otherwise we employ Remark 2 to get us to the case where it is. These arguments establish Harris recurrence of the chain induced by STL with respect to branch lengths, completing the proof. □

This establishes the suitability of STL for use as the sole phylogenetic transition kernel in an MCMC analysis.

# Appendix B: Adaptation scheme

The efficiency of $\hat{\mu}_g$ as an estimator depends crucially on the proposal-generating distribution $Q_\omega(\cdot, \cdot)$, which in turn depends on the indexing parameter $\omega$. In general, $\omega$ can be understood as the *width* of the proposal; if $\omega$ is too small, consecutive states will be highly correlated, and the chain will not mix well. On the other hand, if $\omega$ is too large, proposed values are likely to have low density under the target and hence get rejected. Ideally, one would want to set $\omega$ to an optimal value $\omega^\star$ that maximises the efficiency of the Markov chain, as measured by, say, the effective sample size (ESS). In particular, we would like to find $\omega^\star$ such that the acceptance probability $\alpha$ is at its optimal value, $\alpha^\star$. Theoretical analyses of a host of MCMC algorithms for a broad class of target distributions have shown that $\alpha^\star \approx 0.234$ (0.44 for one-dimensional targets) for random-walk Metropolis (Roberts et al., 1997, 2001), 0.574 for the Metropolis-adjusted Langevin algorithm (MALA) (Roberts et al., 2001) and 0.651 for Hamiltonian Monte Carlo (HMC) (Beskos et al., 2013).

It is convenient to represent the accept-reject mechanism as a binary-valued process with probability $\alpha_\omega$. In particular, we can write (Andrieu and Thoms, 2008):

$$\bar{\alpha}_\omega := \int_{\mathcal{X}\times\mathcal{X}} \alpha_\omega(x,y)\pi_d(x)q_\omega(x,y)dxdy.$$

Recall that in parallel to the chain $\{Z_i\}$ we have a chain of proposed values $\{Y_i\}$. We can formulate the problem of finding $\bar{\alpha}_\omega = \alpha^\star$ as an stochastic approximation problem, more specifically, we can can write (Andrieu and Thoms, 2008, eq. 17):

$$h(\omega) := E_\omega[H(\omega, Z_0, Y_1, Z_1, \ldots)] = 0,$$

where

$$H(\omega, Z_0, Y_1, Z_1, \ldots) := \min\left[1, \frac{\pi_d(Y_1)q_\omega(Y_1, Z_0)}{\pi_d(Z_0)q_\omega(Z_0, Y_1)}\right] - \alpha^\star.$$

This is the so-called **coerced** acceptance probability case, which is implemented in BEAST X (v1.10.5) (Baele et al., 2025). It is equivalent to finding the zeroes of $h(\omega) = \bar{\alpha}_\omega - \alpha^\star$ (Andrieu and Thoms, 2008).

We shall follow Garthwaite et al. (2016) and assume that the acceptance probability $\bar{\alpha}_\omega$ is a monotonically-decreasing function of the scale parameter. The Robbins-Monro algorithm (Robbins and Monro, 1951) is a popular method for solving the zero-finding problem and consists of creating a positive, non-increasing sequence $\{\omega_i\}$, $\omega_i : \Omega \times \mathcal{X} \to \Omega$ via an update of the form (Andrieu and Thoms, 2008, eq. 21)

$$\omega_{n+1} = \omega_n + \gamma_{n+1} h(\omega), \tag{7}$$

subject to the conditions that $\sum_{n=0}^{\infty} \gamma_n = \infty$ and $\sum_{n=0}^{\infty} \gamma_n^2 < \infty$. One way to attain this is to choose $\gamma_n = O(n^{-c})$ for $1/2 < c \leq 1$ (Atchadé et al., 2005). In practice, we need to replace $h(\omega)$ with an estimate, for instance

$$\widehat{h}(\omega) := \widehat{\alpha(\omega)}_n - \alpha^\star,$$
$$\widehat{\alpha(\omega)}_n := \sum_{j=C_0}^{n} \alpha_{\omega_j}(Z_j, Y_j),$$

where $C_0$ is an integer constant chosen so as to avoid transient effects from the initial states of the chain. In BEAST, the Robbins-Monro update is of the form[2]

$$\omega_{n+1} = \omega_n + \frac{1}{f(n)+1}\left(\alpha_{\omega_n}(Z_n, Y_n) - \alpha^\star\right) \tag{8}$$

where $f(x) = x$, $f(x) = \log(x)$ or $f(x) = \sqrt{x}$.

[2]Note that in BEAST X the acceptance rate estimate is **not** smoothed over the Markov chain.